\documentclass[10pt,aps,fleqn,superscriptaddress,notitlepage,nofootinbib,preprintnumbers,nobalancelastpage]{revtex4-1}
\pdfoutput=1
\usepackage{amsmath,amssymb,pstricks,graphicx,xspace,subfigure,todonotes}

\newcommand{\Pythia}{P\protect\scalebox{0.8}{YTHIA}\xspace}
\newcommand{\Sherpa}{S\protect\scalebox{0.8}{HERPA}\xspace}

\newcommand{\Dire}{D\protect\scalebox{0.8}{IRE}\xspace}
\newcommand{\Li}{{\rm Li}\xspace}
\newcommand{\mc}[1]{\mathcal{#1}}

\usepackage{hyperref}
\hypersetup{
  pdfauthor={Stefan Hoeche, Frank Krauss, Stefan Prestel},
  pdftitle={Implementing NLO DGLAP evolution in Parton Showers},
  pdfkeywords={QCD, Parton Shower, NLO}
}
\begin{document}
\preprint{SLAC-PUB-16965}
\preprint{FERMILAB-PUB-17-134-T}
\preprint{IPPP/17/34}
\preprint{DCPT/17/68}
\preprint{MCNET-17-06}
\title{Implementing NLO DGLAP evolution in Parton Showers}
\author{Stefan~H{\"o}che}
\affiliation{SLAC National Accelerator Laboratory,
  Menlo Park, CA, 94025, USA}
\author{Frank~Krauss}
\affiliation{Institute for Particle Physics Phenomenology,
  Durham University, Durham, DH1 3LE, UK}
\author{Stefan~Prestel}
\affiliation{Fermi National Accelerator Laboratory,
  Batavia, IL, 60510-0500, USA}
\begin{abstract}
  We present a parton shower which implements the DGLAP evolution of parton
  densities and fragmentation functions at next-to-leading order precision
  up to effects stemming from local four-momentum conservation.
  The Monte-Carlo simulation is based on including next-to-leading order
  collinear splitting functions in an existing parton shower and combining
  their soft enhanced contributions with the corresponding terms at leading order.
  Soft double counting is avoided by matching to the soft eikonal.
  Example results from two independent realizations of the algorithm,
  implemented in the two event generation frameworks \Pythia and \Sherpa,
  illustrate the improved precision of the new formalism.
\end{abstract}
\maketitle

\section{Introduction}
Parton shower algorithms describing QCD and QED multiple radiation have been
a central ingredient of simulation programs for particle physics experiments
at the energy frontier~\cite{Buckley:2011ms}.  
After their inception about three decades ago, where the focus was on QCD
radiation in the final state~\cite{Fox:1979ag}, efficient algorithms for
initial state radiation were developed~\cite{Sjostrand:1985xi}, which amount
to evolution back in ``time'' from the hard scattering to the incoming beam
hadrons.  The study of quantum interference effects
in successive emissions led to the notion of QCD coherence in parton
evolution~\cite{Azimov:1984np,Azimov:1986sf}, and angular ordering was
identified as a convenient scheme that incorporates such
effects~\cite{Webber:1983if,Marchesini:1983bm,Marchesini:1987cf}.  As an
alternative scheme, the color dipole
model~\cite{Gustafson:1987rq,Lonnblad:1992tz} includes QCD coherence in
a natural way. Matrix-element corrections have been investigated as
a source of coherence~\cite{Bengtsson:1986et,Bengtsson:1986hr}.

After about a decade of work on matching~\cite{Frixione:2002ik,
  Hoeche:2011fd,Hamilton:2012rf,Hamilton:2013fea,Hoeche:2014aia,Hoche:2014dla}
and merging~\cite{Catani:2001cc,Lonnblad:2001iq,Krauss:2002up,Mangano:2006rw,
  Alwall:2007fs,Hoeche:2009rj,Hamilton:2010wh,Hoeche:2012yf,Platzer:2012bs,
  Lonnblad:2012ix,Lonnblad:2012ng,Frederix:2012ps} algorithms, the necessity of increased
control over the parton shower for a more seamless combination with fixed-order
calculations at higher orders triggered a resurgence of interest in improving
parton-shower algorithms themselves.
As a consequence, new parton showers~\cite{Sjostrand:2004ef,Schumann:2007mg,
  Dinsdale:2007mf,Winter:2007ye,Giele:2007di,Platzer:2009jq,Ritzmann:2012ca,
  Hoche:2015sya},
have been constructed that are based on ordering subsequent emissions in
transverse momenta, 
and there were also new constructions with improved and generalized angular
ordering parameters~\cite{Gieseke:2003rz}.
The possibility of including next-to-leading order corrections into parton showers
was explored over three decades ago~\cite{Kato:1986sg,Kato:1988ii,Kato:1990as,Kato:1991fs},
and it was revisited recently in a different framework~\cite{Hartgring:2013jma,Li:2016yez}.
Next-to-leading order corrections to a single final-state gluon emission off a $q\bar{q}$
dipole have been presented in \cite{Hartgring:2013jma} as a first higher-order extension
of the antenna shower formalism. How this approach maps onto NLO DGLAP evolution
was briefly addressed in \cite{Li:2016yez}, which furthermore introduced final-state
double-gluon radiation into this formalism.
In addition to this, NLO splitting functions have been recomputed using a
novel regularization scheme~\cite{Jadach:2011kc,Gituliar:2014eba}, with the
aim to improve parton-shower simulations. The dependence of NLO matching
terms on the parton-shower evolution variable has also been investigated~\cite{Jadach:2016zgk}.

This publication is dedicated to the construction of a parton shower that
implements the next-to-leading order (NLO) DGLAP equations up to momentum
conserving effects. We employ the non-flavor changing NLO splitting functions
in the $\overline{\rm MS}$ scheme in their integrated form~\cite{Curci:1980uw,
  Furmanski:1980cm,Floratos:1980hk,Floratos:1980hm,Heinrich:1997kv,Ellis:1996nn},
and we include the flavor-changing NLO splitting kernels fully differentially
using the method presented in~\cite{Hoche:2017iem}.  We identify
the contribution to the NLO splitting functions which is already included in
the leading-order (LO) realization of the parton-shower, and correspondingly
subtract it from the NLO splitting function.
This term is given by the two-loop cusp anomalous dimension, which is usually
included at LO using the CMW scheme~\cite{Catani:1990rr}.  After its subtraction,
the remaining splitting function is purely collinear, and no double-counting
arises upon implementing it as a higher-order correction to the existing splitting
kernels of the parton shower.
However, the NLO parts of the splitting functions are negative in large parts
of the phase space which presents a technical challenge.  We overcome this problem
through the  weighting algorithm first proposed
in~\cite{Hoeche:2009xc,Lonnblad:2012hz}.
Our approach can be considered as a first step towards a fully next-to-leading
order accurate parton shower and acts as a baseline for further development.
Future projects will need to address the leading-color approximation and
the simulation of soft emissions beyond the leading order. The Monte-Carlo
techniques developed here are expected to become useful in this context as well.
A clear phenomenological benefit of the present implementation is that consistency
between the parton shower and NLO PDF evolution is achieved for the very first time.

The outline of this paper is as follows. Section~\ref{sec:ps_formalism}
introduces the parton-shower formalism at leading order and establishes the
connection to the DGLAP equation in order to identify the correct treatment
of the final-state Sudakov factor.  Section~\ref{sec:implementation} outlines
the specific implementation in the \Dire parton showers~\cite{Hoche:2015sya}.
First results and applications are presented in Sec.~\ref{sec:results}.
Section~\ref{sec:conclusions} contains our conclusions.

\section{Extension of the parton-shower formalism}
\label{sec:ps_formalism}
In this section we will highlight the correspondence between the parton shower
formalism and the analytic structure of the DGLAP evolution equations~\cite{
  Gribov:1972ri,*Dokshitzer:1977sg,*Altarelli:1977zs},
on which the parton shower is based. We will thereby focus on the refinements
needed in order to realize NLO accurate parton evolution. This includes
the implementation of the complete set of splitting kernels at $\mc{O}(\alpha_s^2)$,
a subset of which are the flavor-changing kernels discussed in~\cite{Hoche:2017iem}.
Another significant change concerns the implementation of symmetry factors. 
In the computation of the NLO splitting functions~\cite{Curci:1980uw,
  Furmanski:1980cm,Floratos:1980hk,Floratos:1980hm,Heinrich:1997kv,Ellis:1996nn},
it is assumed that a certain final-state parton is identified, while the parton shower
treats all particles democratically. If the full set of splitting functions
-- both at leading and at next-to-leading order -- is implemented naively,
the emission probability will thus be overestimated. At leading-order the problem
can be solved by adding simple symmetry factors. At next-to-leading order the
solution will be different, as the splitting functions include contributions
from three-particle final states that have been integrated out. We will show
in the following how this problem can be approached~\cite{Hoche:2017iem}.

\subsection{Unconstrained evolution with identified final-states}
The DGLAP equations are schematically identical for initial and final state.
However, their implementation in parton-shower programs differs between the two,
owing to the fact that Monte-Carlo simulations are typically performed for
inclusive final states.
The inclusive evolution equations for the fragmentation functions $D_a(x,Q^2)$
for parton of type $a$ to fragment into a hadron read
\begin{equation}\label{eq:pdf_evolution}
  \frac{{\rm d}\,xD_{a}(x,t)}{{\rm d}\ln t}=
  \sum_{b=q,g}\int_0^1{\rm d}\tau\int_0^1{\rm d} z\,\frac{\alpha_s}{2\pi}
  \big[zP_{ab}(z)\big]_+\,\tau D_{b}(\tau,t)\,\delta(x-\tau z)\;,
\end{equation}
where the $P_{ab}$ are the unregularized DGLAP evolution kernels, and the
plus prescription is defined to enforce the momentum and flavor sum rules:
\begin{equation}\label{eq:sf_regularization}
  \big[zP_{ab}(z)\big]_+=\lim\limits_{\varepsilon\to 0}
  \bigg[zP_{ab}(z)\,\Theta(1-z-\varepsilon)
  -\delta_{ab}\sum_{c\in\{q,g\}}
  \frac{\Theta(z-1+\varepsilon)}{\varepsilon}
  \int_0^{1-\varepsilon}{\rm d}\zeta\,\zeta\,P_{ac}(\zeta)\bigg]\;.
\end{equation}
For finite $\varepsilon$, the endpoint subtraction in Eq.~\eqref{eq:sf_regularization} 
can be interpreted as the approximate virtual plus unresolved real corrections, 
which are included in the parton shower because the Monte-Carlo algorithm
naturally implements a unitarity constraint~\cite{Jadach:2003bu}.  The precise value 
of $\varepsilon$ in this case depends on the infrared cutoff on the evolution variable,
and is determined by local four-momentum conservation in the parton branching process.
For $0<\varepsilon\ll 1$, Eq.~\eqref{eq:pdf_evolution} changes to
\begin{equation}\label{eq:pdf_evolution_constrained}
  \frac{1}{D_{a}(x,t)}\,\frac{{\rm d} D_{a}(x,t)}{{\rm d}\ln t}=
  -\sum_{c=q,g}\int_0^{1-\varepsilon}{\rm d}\zeta\,\zeta\,\frac{\alpha_s}{2\pi}P_{ac}(\zeta)\,
  +\sum_{b=q,g}\int_x^{1-\varepsilon}\frac{{\rm d} z}{z}\,
  \frac{\alpha_s}{2\pi}\,P_{ab}(z)\,\frac{D_{b}(x/z,t)}{D_{a}(x,t)}\;.
\end{equation}
Using the Sudakov form factor
\begin{equation}\label{eq:sudakov}
  \Delta_a(t_0,t)=\exp\bigg\{-\int_{t_0}^{t}\frac{{\rm d} \bar{t}}{\bar{t}}
  \sum_{c=q,g} \int_0^{1-\varepsilon}{\rm d}\zeta\,\zeta\,\frac{\alpha_s}{2\pi}P_{ac}(\zeta)\bigg\}
\end{equation}
the generating function for splittings of parton $a$ is defined as
\begin{equation}\label{eq:def_updf}
  \mc{D}_a(x,t,\mu^2)=D_a(x,t)\Delta_a(t,\mu^2)\,.
\end{equation}
Equation~\eqref{eq:pdf_evolution_constrained} can now be written in the simple form
\begin{equation}\label{eq:pdf_evolution_constrained_2}
  \frac{{\rm d}\ln\mc{D}_a(x,t,\mu^2)}{{\rm d}\ln t}
  =\sum_{b=q,g}\int_x^{1-\varepsilon}\frac{{\rm d} z}{z}\,
  \frac{\alpha_s}{2\pi}\,P_{ab}(z)\,\frac{D_{b}(x/z,t)}{D_{a}(x,t)}\;.
\end{equation}
The generalization to an $n$-parton state, $\vec{a}=\{a_1,\ldots,a_n\}$,
with jets and incoming hadrons resolved at scale $t$ can be made in terms
of parton distribution functions (PDFs) $f$, and fragmenting jet functions,
$\mc{G}$~\cite{Procura:2009vm,Jain:2011xz}.
If we define the generating function for this state as
$\mc{F}_{\vec{a}}(\vec{x},t,\mu^2)$, we can formulate
its evolution equation in terms of a sum of the right hand side
of Eq.~\eqref{eq:pdf_evolution_constrained_2}, where each term in the sum
corresponds to a resolved jet in the final state or a hadron in the initial state.
This equation can be solved using Markovian Monte-Carlo techniques in the form of
a parton shower~\cite{Buckley:2011ms}. In most cases, however, parton showers
implement final-state branchings in unconstrained evolution, which means that
final-state hadrons are not resolved. We can use Eq.~\eqref{eq:pdf_evolution_constrained}
(which also applies to $\mc{G}$~\cite{Procura:2009vm,Jain:2011xz}),
to write the corresponding differential decay probability for such an evolution as
\begin{equation}\label{eq:pdf_evolution_constrained_4}
  \begin{split}
    \frac{{\rm d}}{{\rm d}\ln t}\ln\bigg(
    \frac{\mc{F}_{\vec{a}}(\vec{x},t,\mu^2)}{\prod_{j\in\rm FS}\mc{G}_{a_j}(x_j,t)}\bigg)
  =&\sum_{i\in{\rm IS}}\sum_{b=q,g}\int_{x_i}^{1-\varepsilon}\frac{{\rm d} z}{z}\,
  \frac{\alpha_s}{2\pi}\,P_{ba_i}(z)\,\frac{f_{b}(x_i/z,t)}{f_{a_i}(x_i,t)}
  +\sum_{j\in{\rm FS}}\sum_{b=q,g}\int_0^{1-\varepsilon}{\rm d} z\,z\,
  \frac{\alpha_s}{2\pi}\,P_{a_jb}(z)\;.
  \end{split}
\end{equation}
As highlighted in~\cite{Jadach:2003bu}, it is necessary to use the Sudakov factor,
Eq.~\eqref{eq:sudakov}, in final-state parton showers beyond the leading order.
At the leading order, the factor $\zeta$ in Eq.~\eqref{eq:sudakov} simply replaces
the commonly used symmetry factor for $g\to g$ splitting and it also accounts for the
proper counting of the number of active flavors. However, at the next-to-leading order
it becomes an identifier for the parton that undergoes evolution, which is essential
in order to obtain the correct anomalous dimensions upon integration of the NLO
DGLAP evolution kernels. We will thus define the final-state Sudakov factor
in our implementation according to Eq.~\eqref{eq:sudakov}.

\subsection{Splitting functions}
The crucial ingredient of NLO DGLAP evolution are the $\mc{O}(\alpha_s^2)$ corrections
to the evolution kernels. These corrections depend on the scheme in which
PDFs and fragmentation functions are renormalized. We will work in the
$\overline{\rm MS}$ scheme, which allows us to use the results of \cite{Curci:1980uw}.
Technical challenges in the implementation of the splitting functions in the parton shower
include the overlap with the CMW scheme for setting the renormalization scale commonly
used in leading-order parton showers~\cite{Catani:1990rr} as well as the fact
that the evolution kernels are negative in large parts of the accessible phase space.
We will discuss these problems in a general context in the following and give more details
on the implementation in the Dire parton showers in Sec.~\ref{sec:implementation}.

At $\mc{O}(\alpha_s)$, the unregularized DGLAP splitting functions are
\begin{equation}
  \begin{split}
    P_{qq}^{(0)}(z)=&\;C_F\bigg[\frac{1+z^2}{1-z}\bigg]\;,
    \qquad P_{gq}^{(0)}(z)=T_R\Big[1-2z(1-z)\Big]\;,\\
    P_{gg}^{(0)}(z)=&\;2C_A\bigg[\frac{z}{1-z}+\frac{1-z}{z}+z(1-z)\bigg]\;.
  \end{split}
\end{equation}
At $\mc{O}(\alpha_s^2)$, the quark splitting functions are typically written
in terms of singlet (S) and non-singlet (V) components as
\begin{equation}\label{eq:ap_kernels_1_qdef}
  \begin{split}
    P_{qq}^{(1)}(z)=&\;p_{qq}^{V(1)}(z)+p_{qq}^{S(1)}(z)\;,
    \qquad &P_{q\bar{q}}^{(1)}(z)=&\;p_{q\bar{q}}^{V(1)}(z)+p_{qq}^{S(1)}(z)\;,
    \qquad &P_{qq'}^{(1)}(z)=&\;P_{q\bar{q}'}^{(1)}(z)=p_{qq}^{S(1)}(z)\;.
  \end{split}
\end{equation}
In the timelike case, their components are
\begin{equation}\label{eq:nlo_sf_example}
  \begin{split}
    p_{qq}^{S(1)}(z)=&\;C_FT_F\left[\,(1+x)\log^2 x
      -\left(\frac{8}{3}z^2+9z+5\right)\log z
      +\frac{56}{9}z^2+4z-8-\frac{20}{9z}\,\right]\;,\\
    p_{qq}^{V(1)}(z)=&\;
    p^{(0)}_{qq}(z)\left[\Big(\beta_0\log z+\Gamma^{(2)}\Big)
      +2C_F\log z\left(\log\frac{1-z}{z}+\frac{3}{4}\right)+
      \frac{C_A}{2}\log^2 z\right]
    -\frac{4}{3}C_F T_F(1-z)\\
    &-C_F^2\,\left[\left(\frac{7}{2}+\frac{3}{2}z\right)\log z
      -\frac{1}{2}(1+z)\log^2z+5(1-z)\right]
    +C_FC_A\,\left[(1+z)\log z+\frac{20}{3}(1-z)\right]\;.
  \end{split}
\end{equation}

The flavor-changing splitting kernels, $P_{qq'}^{(1)}$ and $P_{q\bar{q}}^{(1)}$
first appear at $\mc{O}(\alpha_s^2)$. They are new channels contributing to the
real-emission corrections to $P_{qq}^{(0)}$. In order to account for their
more involved flavor structure, we must simulate them fully differentially
in the $1\to 3$ phase space. To this end, we use the method presented
in~\cite{Hoche:2017iem}.  All other splitting functions have an analogous
$1\to 2$ topology and are implemented using this topology.

Several new structures appear in the next-to-leading order splitting functions,
which require a modification of the branching algorithm used at the leading-order.
Firstly, the NLO splitting functions may exhibit new types of apparent singularities,
like the term $-20/9\,C_FT_F/z$ contributing to $p_{qq}^{S(1)}$.  Such terms are
regulated by the symmetry factor in Eq.~\eqref{eq:sudakov}, which highlights
again that without the correct definition of the Sudakov factor one cannot construct
a meaningful Monte-Carlo implementation, as the resulting integrals would have
unphysical divergences.

In addition, $p_{qq}^{V(1)}$ and $P_{gg}^{(1)}$ include the two-loop cusp anomalous
dimension, given by~\cite{Catani:1990rr}
\begin{equation}
  \Gamma^{(2)}=\left(\frac{67}{18}-\frac{\pi^2}{6}\right)C_A-\frac{10}{9}\,T_F.
\end{equation}
This term is routinely included in standard parton-shower Monte Carlo simulation,
typically through a redefinition of the scale at which the strong coupling is
evaluated~\cite{Catani:1990rr}. It must therefore be subtracted from $p_{qq}^{V(1)}$
and $P_{gg}^{(1)}$ before these splitting functions can be included.  After the
subtraction, no soft enhanced terms remain, and the result is a purely collinear
splitting function. This is important to avoid double counting of singular limits
in the parton shower~\cite{Marchesini:1987cf}.
Furthermore, $p_{qq}^{V(1)}$ and $P_{gg}^{(1)}$ also contain a term originating from
the renormalization of the strong coupling constant, which is given by
the leading-order splitting function times $\beta_0\log z$, where
\begin{equation}\label{eq:beta_coefficients}
  \beta_0=\frac{11}{6}\,C_A-\frac{2}{3}\,T_F\;.
\end{equation}
The leading contribution from this term upon integration over $z$ is generated
in combination with the soft factor $2/(1-z)$ of the leading-order splitting function,
and gives a contribution $-\beta_0\pi^2/3$ to the collinear anomalous dimension.

\section{Implementation in the Dire parton shower}
\label{sec:implementation}
Our numerical simulations are based on the \Dire parton shower, presented
in~\cite{Hoche:2015sya}.  This section first presents a brief overview of the
model as implemented at leading order, before moving to the modifications needed
for an implementation of the next-to-leading order contributions.

\subsection{Parton-shower model at leading order}
\label{sec:dire_lo}
The evolution and splitting parameters $\kappa$ and $z_j$ used in \Dire
for splittings of a combined parton $ij$ to partons $i$ and $j$ in the presence
of a spectator $k$ are given by
\begin{eqnarray}
  \kappa_{j,ik}^2 = \frac{4\,(p_ip_j)\,(p_jp_k)}{Q^4}
  \quad\text{and}\quad
  z_j = \frac{2\,p_jp_k}{Q^2}\;.
\end{eqnarray}
In this context, $Q^2$ plays the role of the maximally attainable momentum
squared, which is defined as $Q^2_{\rm FF}=2(p_i+p_j)p_k+2p_ip_j$ for final-state
splittings with final-state spectator, $Q^2_{\rm FI}=Q^2_{\rm IF}=2(p_i+p_j)p_k$
for final(initial)-state splittings with initial(final)-state spectator,
and $Q^2_{\rm II}=2p_ip_k$ for initial-state splittings with initial-state
spectator. The splitting functions for initial-state branchings are given
by the modified DGLAP splitting functions~\cite{Hoche:2015sya}
\begin{equation}\label{eq:dire_lo_sf}
  \begin{split}
  P^{(0)}_{qq}(z,\,\kappa^2) =&\;
  2C_F\left[\frac{1-z}{(1-z)^2+\kappa^2}-\frac{1+z}{2}\right]\;,
  \quad
  &P^{(0)}_{qg}(z,\,\kappa^2) =&\;2C_F\left[\frac{z}{z^2+\kappa^2}-\frac{2-z}{2}\right]\\
  P^{(0)}_{gg}(z,\,\kappa^2) =&\;
  2C_A\left[\frac{1-z}{(1-z)^2+\kappa^2}+\frac{z}{z^2+\kappa^2}-2+z(1-z)\right]\;,
  \quad
  &P^{(0)}_{gq}(z,\,\kappa^2) =&\;T_R\left[z^2+(1-z)^2\right]\,.
  \end{split}
\end{equation}
where $z=1-z_j$. It is interesting to note that the dimensionless quantity
$\kappa^2$ plays the role of the IR regulator in the very same fashion as
the principal value regulator $\delta^2$ introduced in Eq.~(3.13)
of~\cite{Curci:1980uw}. In our algorithm, $\kappa$ has a physical interpretation,
as the scaled transverse momentum in the soft limit.  As such, it also sets the
renormalization and factorization scale through $\mu_{R/F}^2=\kappa^2 Q^2$.
For final-state branchings, the matching to the differential cross section
in the soft limit requires the replacement
\begin{equation}\label{eq:dire_gg_soft_match}
  P^{(0)}_{gg}\to P^{s(0)}_{gg}(1-z_j,\kappa_{j,ik}^2)
  +P^{s(0)}_{gg}(1-z_i,\kappa_{i,jk}^2)\;,
\end{equation}
where the $j$-soft part of the splitting function is given by
\begin{equation}
  P^{s(0)}_{gg}(z,\,\kappa^2)=
  2C_A\left[\frac{1-z}{(1-z)^2+\kappa^2}-1+\frac{z(1-z)}{2}\right]\;.
\end{equation}
In a similar fashion we have
\begin{equation}
  P^{(0)}_{qg}\to P^{(0)}_{qg}(1-z_j,\kappa_{i,jk}^2)\;.
\end{equation}
The two terms in Eq.~\eqref{eq:dire_gg_soft_match} correspond to different
color flows in the parton shower. For the first term partons $i$ and $k$
are considered radiators and $j$ is the soft gluon insertion, while for
the second term partons $j$ and $k$ are the radiators and $i$ is
the soft gluon insertion. Therefore, in the first term gluon $j$ is
color-connected to the spectator parton, while in the second term gluon
$i$ is color-connected to the spectator. The two contributions are evolved
using the two different variables $\kappa_{j,ik}^2$ and $\kappa_{i,jk}^2$.
Following standard practice to improve the logarithmic accuracy of the
parton shower, the soft enhanced term of the splitting functions,
Eqs.~\eqref{eq:dire_lo_sf}, is rescaled by
$1+\alpha_s(t)/(2\pi)\,\Gamma^{(2)}$~\cite{Catani:1990rr}. We do not
absorb this rescaling into a redefinition of the strong coupling, as
this would generate higher-logarithmic contributions stemming from
the interaction with the purely collinear parts of the splitting functions.

\subsection{Extension to the next-to-leading order}
\label{sec:dire_nlo}
We now describe the extensions of the \Dire parton shower that are necessary
to construct a simulation which describes the DGLAP evolution of parton
distributions and fragmentation functions at next-to leading order precision.
As an important construction paradigm, we consider contributions at different
orders in the strong coupling as separate evolution kernels, and we restrict
ourselves to the inclusive radiation pattern where possible.  The latter
implies that in general we do not attempt to simulate the emission of an
unordered pair of partons according to the triple collinear splitting functions.
The notable exception to this is the treatment of flavor-changing splitting functions,
where the implementation of a $1\to 2$ rather than a $1\to 3$ transition is
not possible due to local flavor conservation.  The generation of these
contributions is described in detail in~\cite{Hoche:2017iem}\footnote{%
  The contribution from triple collinear splitting functions of type $q\to q'$
  and $q\to\bar{q}$ to the overall NLO corrections is numerically small.
  A more detailed discussion can be found in~\cite{Hoche:2017iem}.}. The main remaining
complication in the implementation of the integrated NLO splitting functions arises
from the fact that they assume negative values in large regions of phase space,
hence we potentially need to generate branchings based on negative ``probabilities''.
To this end we use the method developed in~\cite{Hoeche:2009xc,Hoeche:2011fd,Lonnblad:2012hz}.

We start by formally replacing the leading-order splitting functions of
Eq.~\eqref{eq:dire_lo_sf} with the combined leading-order plus next-to-leading
order evolution kernels.\footnote{For a complete list of the NLO splitting
  functions see App.~\ref{app:coefficient_functions}. Note that we do not
  require the knowledge of $p_{q\bar{q}}^{V(1)}$ in our approach, because
  flavor-changing splittings are generated fully differentially in the
  $1\to 3$ phase space.}
\begin{equation}
  P_{ab}(z,\kappa^2)=P^{(0)}_{ab}(z,\kappa^2)+
  \frac{\alpha_s}{2\pi}\,P^{(1)}_{ab}(z,\kappa^2)\;.
\end{equation}
As described in Sec.~\ref{sec:ps_formalism}, the soft enhanced part of
$P_{qq}^{(1)}$ and $P_{gg}^{(1)}$ matches the term
$\alpha_s/(2\pi)\Gamma^{(2)}\,2C_a/(1-z)$, at leading order, which is
included in the implementation of the leading-order parton shower by
rescaling the soft enhanced part of the splitting functions.  We therefore
subtract this contribution from the NLO splitting kernel and define
\begin{equation}
  P^{(1)}_{ab}(z,\kappa^2)\to P^{(1)}_{ab}(z)
  -\delta_{ab}\,\frac{2C_a}{1-z}\,\Gamma^{(2)}\;.
\end{equation}
In addition, we include  in the soft enhanced part of the leading-order
splitting function the three-loop coefficient $\Gamma^{(3)}$, computed
in~\cite{Becher:2010tm}\footnote{The normalization differs by factor four
  between our notation and that of~\cite{Becher:2010tm}.}
\begin{equation}
  P^{(0)}_{ab}(z,\kappa^2)\to
  P^{(0)}_{ab}(z,\kappa^2)+
  \delta_{ab}\,2C_a\frac{1-z}{(1-z)^2+\kappa^2}\frac{\alpha_s}{2\pi}
  \left[\,\Gamma^{(2)}+\frac{\alpha_s}{2\pi}\Gamma^{(3)}\,\right]\;.
\end{equation}
For final-state gluon evolution this requires the independent modification of
both terms in Eq.~\eqref{eq:dire_gg_soft_match}.

Scale variations can be performed in the \Dire showers by using a method
similar to~\cite{Badger:2016bpw}.  When varying the argument of the strong
coupling, {\it i.e.}\ replacing $\alpha_s(t)\to \alpha_s(c\,t)\,f(c,t)$,
with $c$ a constant, the appropriate counterterm at $\mc{O}(\alpha_s^2)$,
which multiplies the leading-order splitting functions, $P_{ab}^{(0)}$,
reads\footnote{Note that the lowest-order DGLAP kernels, $P_{ab}^{(0)}$,
  are defined at $\mc{O}(\alpha_s)$, and we use a strict order counting.
  The scale variations in our approach are therefore more conservative
  than the ones presented in~\cite{Badger:2016bpw}.}
\begin{equation}\label{eq:asct}
  \begin{split}
    f(c,t)=
    \prod_{i=0}^{n_\text{th}+1}\left[\,1
      -\frac{\alpha_s}{2\pi}\,\beta_0(\bar{t})L
      \,\right]\;,
    \quad\text{where}\quad
    L=\log\frac{t_i}{t_{i-1}},\quad
    \bar{t}=\frac{t_i+t_{i+1}}{2}\;.
  \end{split}
\end{equation}
We use the multiplicative threshold matching described in~\cite{Badger:2016bpw},
as the additive matching generates artificially large deviations in the case of
two-loop and three-loop running of the coupling.
The product in Eq.~\eqref{eq:asct} runs over the number $n_\text{th}$
of parton mass thresholds in the interval $(t,c\cdot t)$ with $t_0=t$,
$t_{n_\text{th}+1}=c\cdot t$ and $t_i$ are the encompassed parton mass
thresholds.  If $c<1$, the ordering is reversed to recover the correct
sign.  $\beta_0(\bar{t})$ is the QCD beta function coefficient, which
depends on the scale $\bar{t}$ through the number of active parton flavors.

\section{Dire predictions}
\label{sec:results}
We have implemented our new algorithms into the \Dire parton showers,
which implies two entirely independent realizations within the general purpose event
generation frameworks \Pythia~\cite{Sjostrand:1985xi,Sjostrand:2014zea}
and \Sherpa~\cite{Gleisberg:2003xi,Gleisberg:2008ta}.
This section presents a first application of our new algorithm to the simulation of the
reactions $e^+e^-\to$hadrons, $pp\to e^+\nu_e$ and $pp\to h$. We compare the magnitude of the
next-to-leading order corrections and the size of their uncertainties to the respective
leading-order predictions. Note that we only quote the renomalization scale uncertainties,
which are the ones that can be expected to decrease when moving from leading to next-to-leading
order evolution. There are of course other uncertainties, for example those related to
the kinematics mapping and the choice of the evolution variable in the parton shower.
However, these effects arise identically both at leading and at next-to-leading order,
and they are therefore not included in the uncertainty bands. In addition, nonperturbative
effects will contribute their own uncertainty, which is somewhat harder to quantify.
However, it is expected that a reduced perturbative uncertainty will lead to a more
realistic extraction of nonperturbative model parameters, and that the uncertianties
on those parameters can therefore be reduced as well.

\begin{figure}[p]
  \centering
  \includegraphics[width=5.75cm]{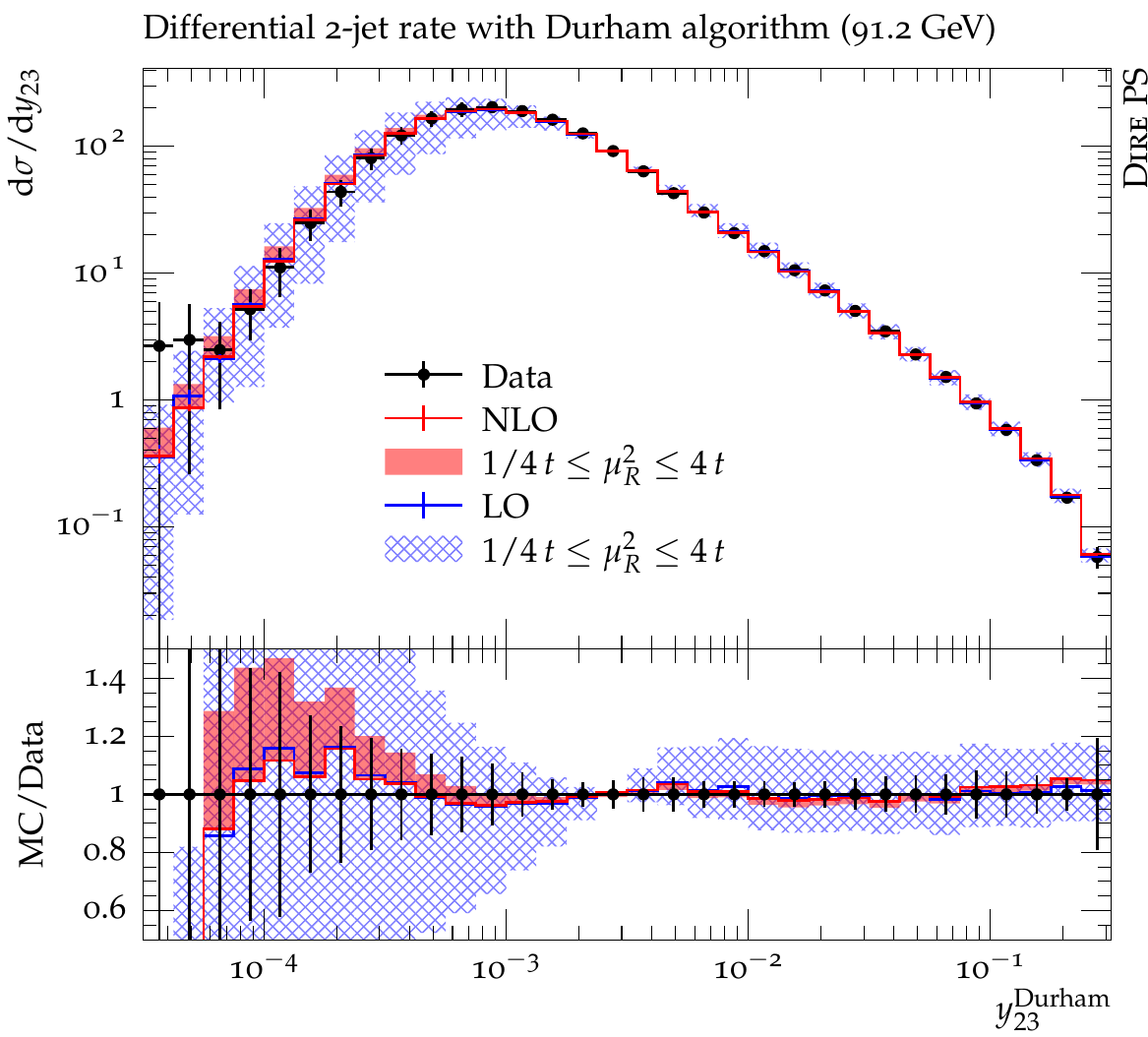}
  \includegraphics[width=5.75cm]{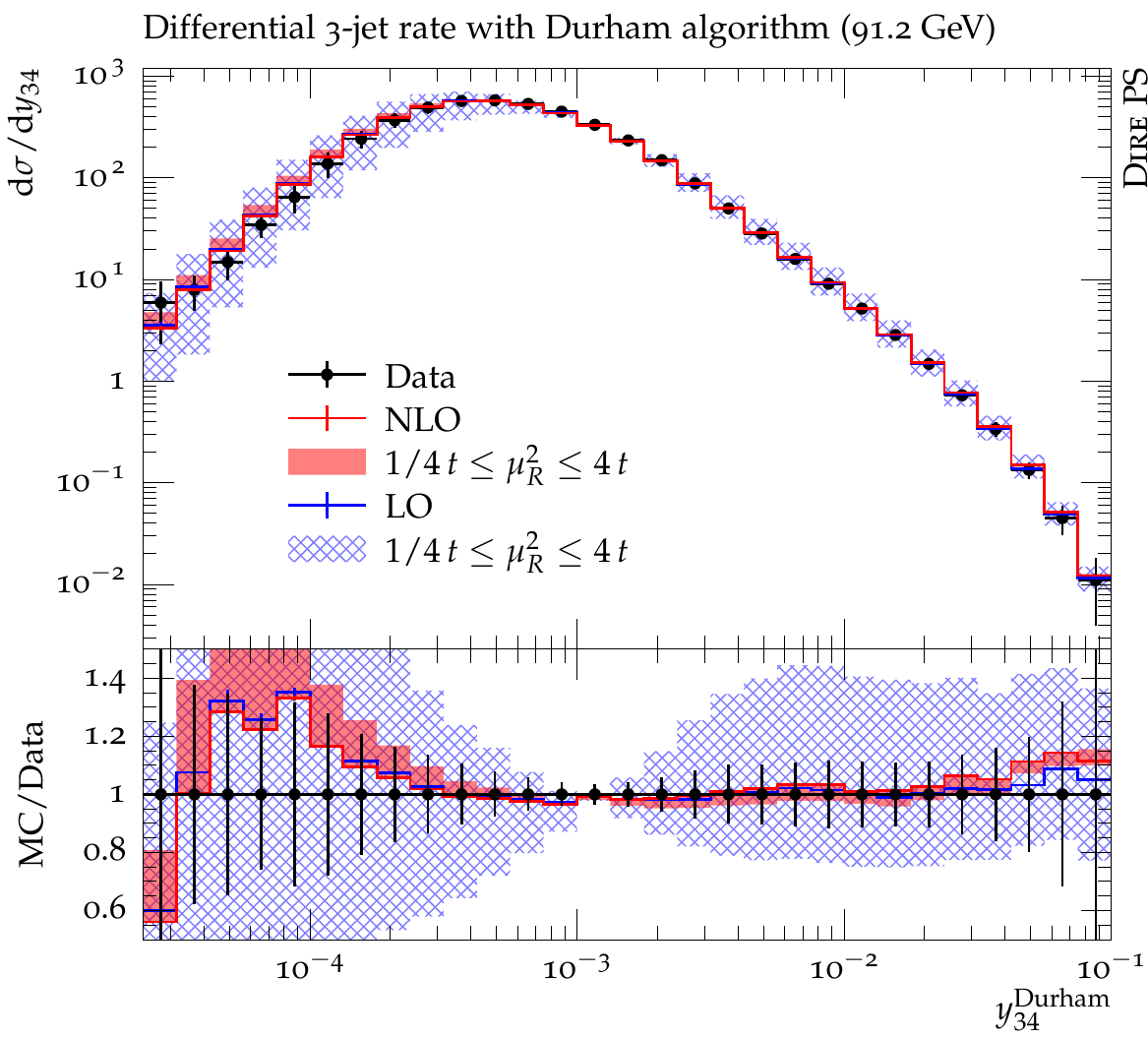}\\
  \includegraphics[width=5.75cm]{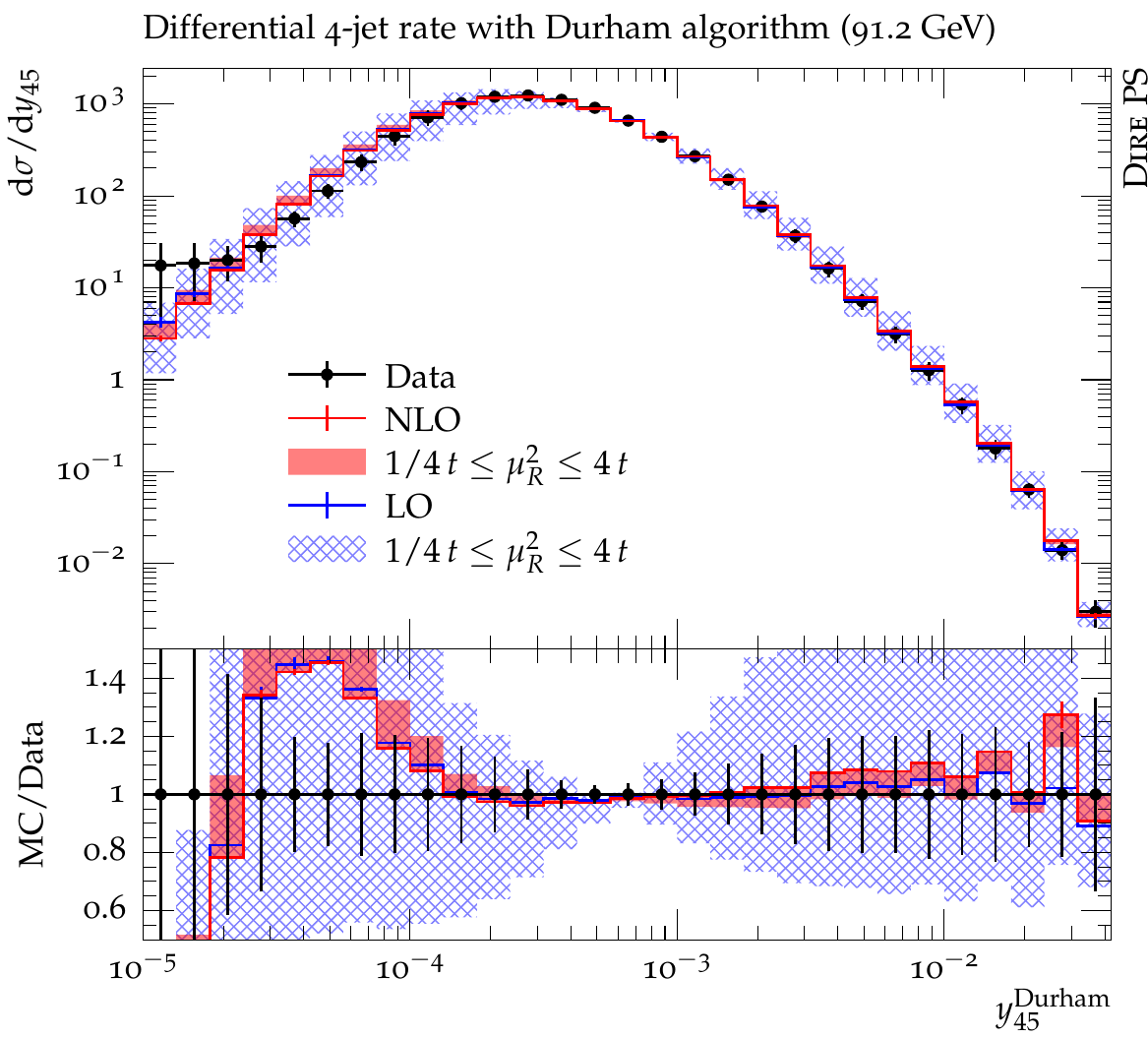}
  \includegraphics[width=5.75cm]{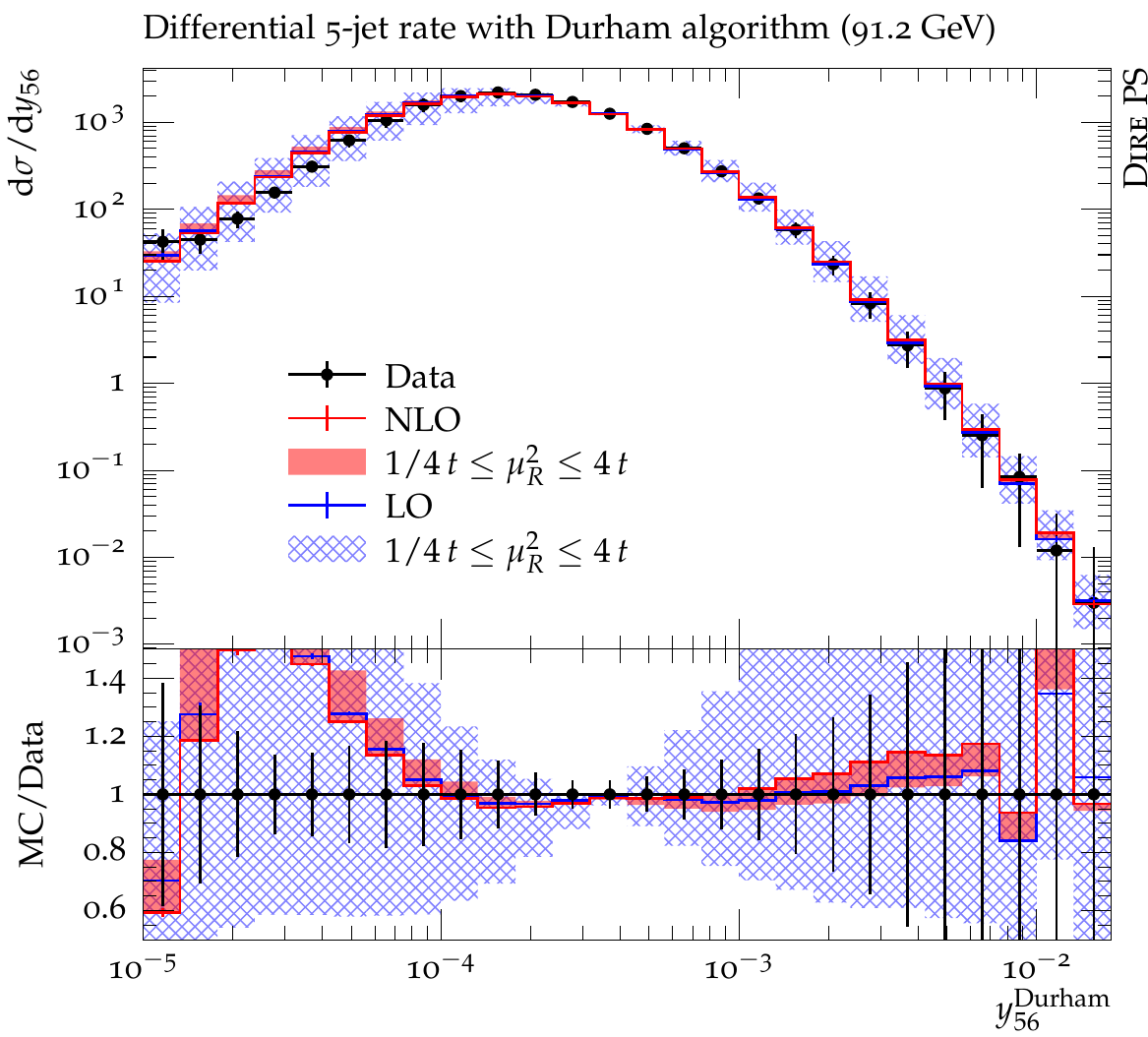}
  \caption{Results for leading and next-to-leading order DGLAP evolution
    in comparison to LEP data from~\cite{Pfeifenschneider:1999rz}.
    \label{fig:lep_jetrates}}
\end{figure}

Figure~\ref{fig:lep_jetrates} shows predictions from our new implementation
compared to leading-order results from the \Dire parton shower for differential 
jet rates in the Durham scheme compared to experimental results from the JADE and OPAL 
collaborations~\cite{Pfeifenschneider:1999rz}. Results have been obtained with
\Dire{}+\Sherpa using the default settings of \Sherpa version 2.2.3. The perturbative region
is to the right of the plots, and $y\sim 2.8\cdot10^{-3}$ corresponds to the $b$-quark mass.
The simulation of nonperturbative effects dominates the predictions below $\sim10^{-4}$.
In the perturbative region, the results are in excellent agreement with the experimental
measurements. The shapes of distributions receive only minor changes compared to
the leading-order result, however, the uncertainties are greatly reduced.

\begin{figure}[p]
  \centering
  \includegraphics[width=5.75cm]{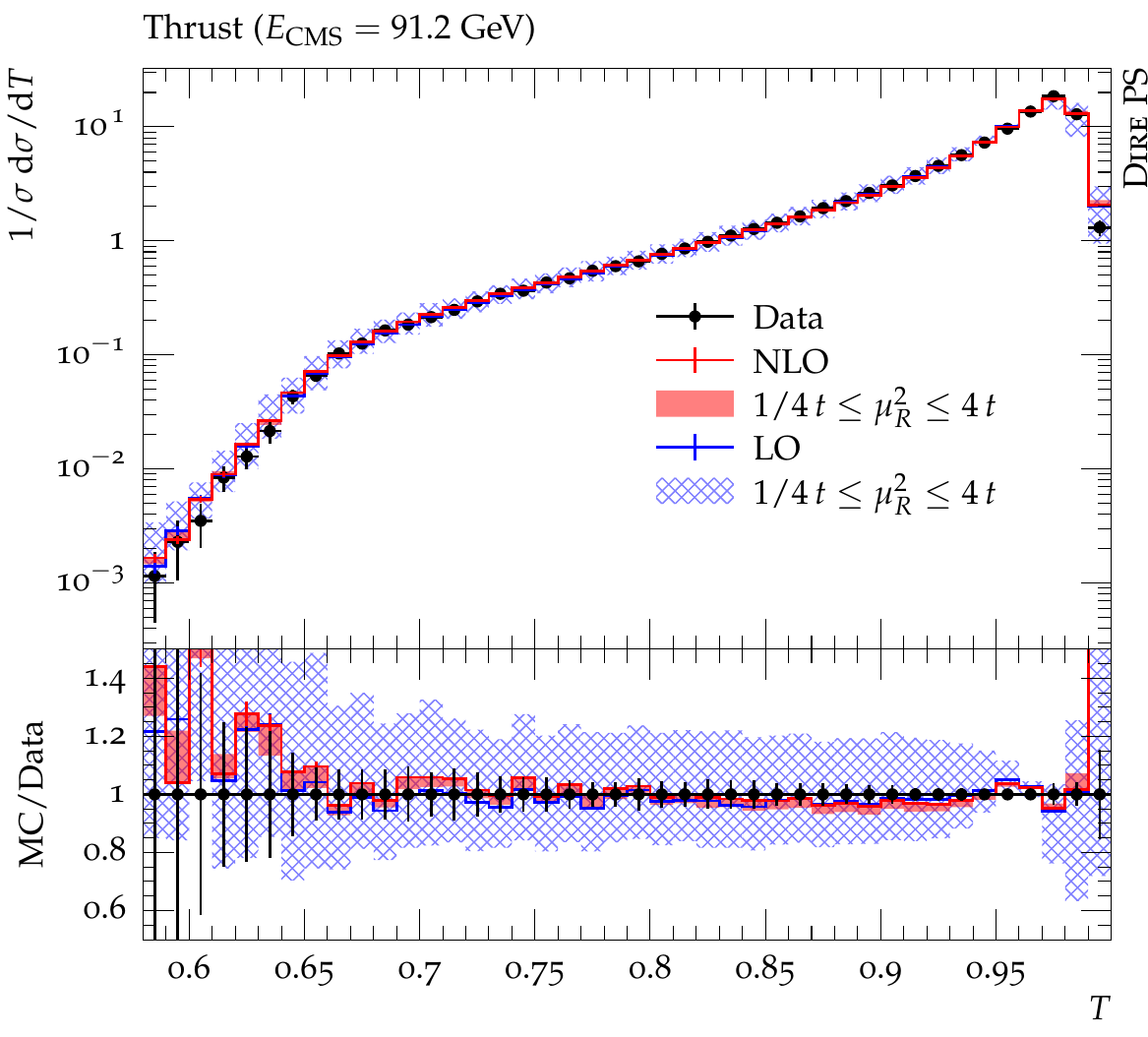}
  \includegraphics[width=5.75cm]{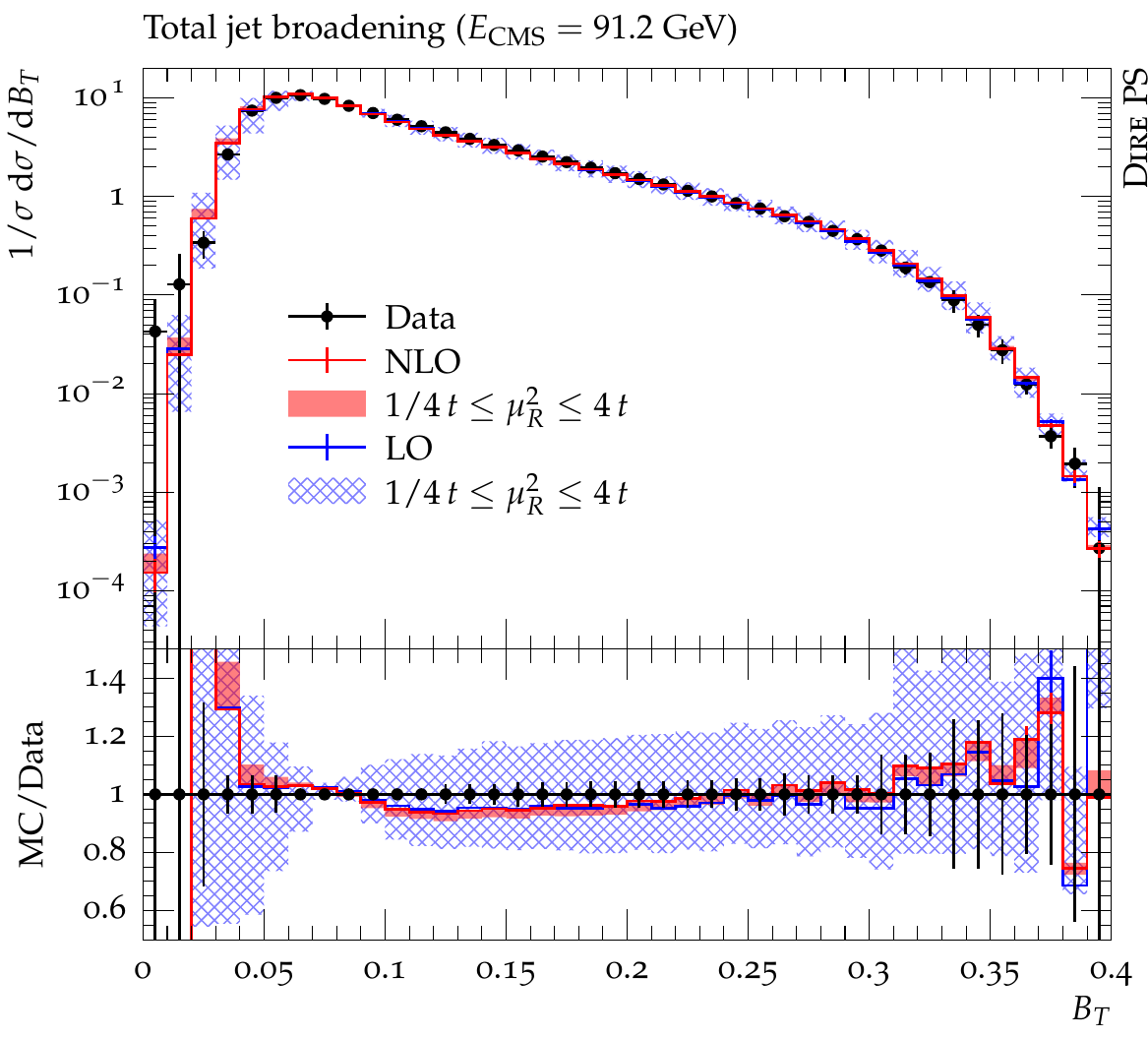}\\
  \includegraphics[width=5.75cm]{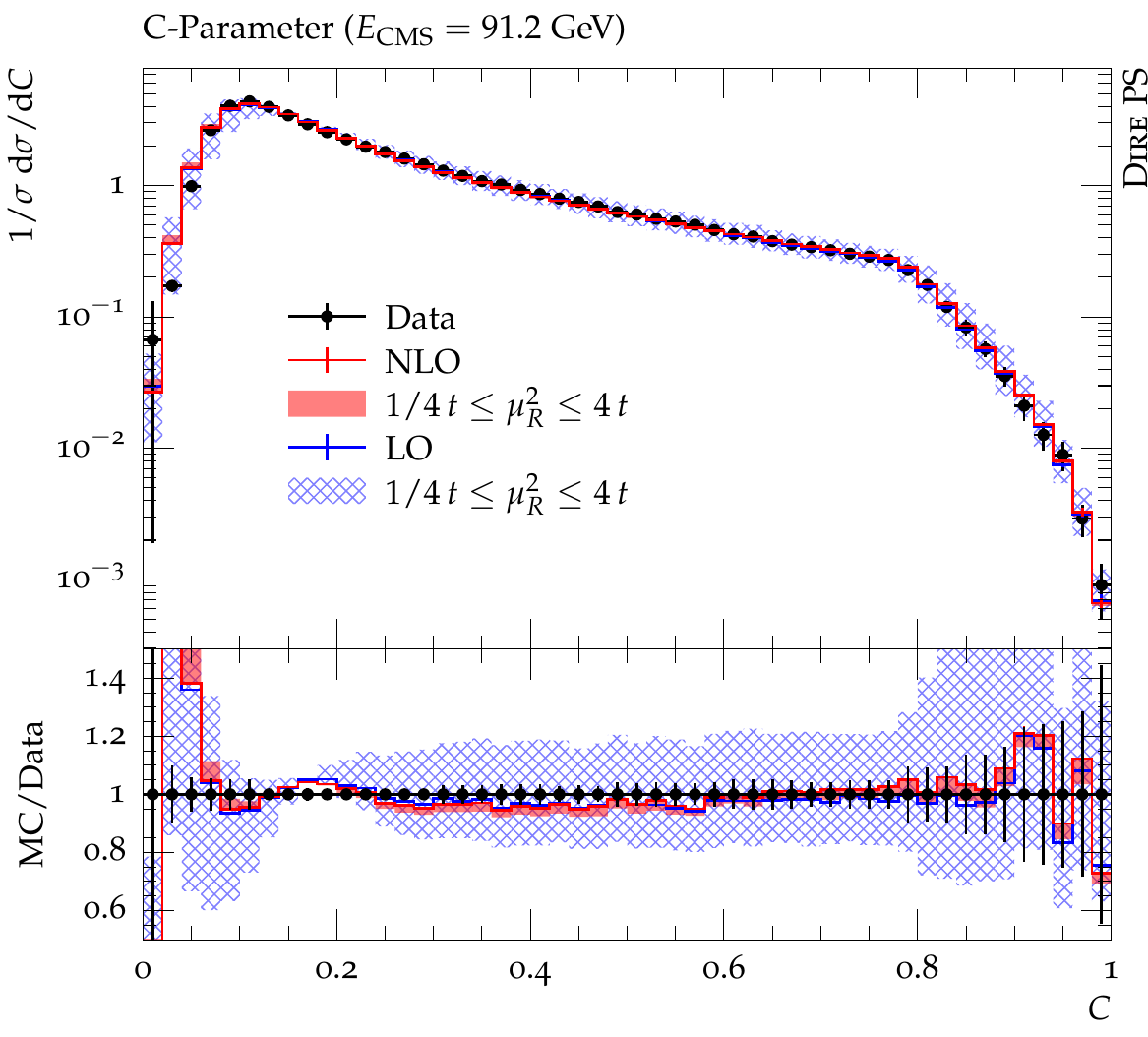}
  \includegraphics[width=5.75cm]{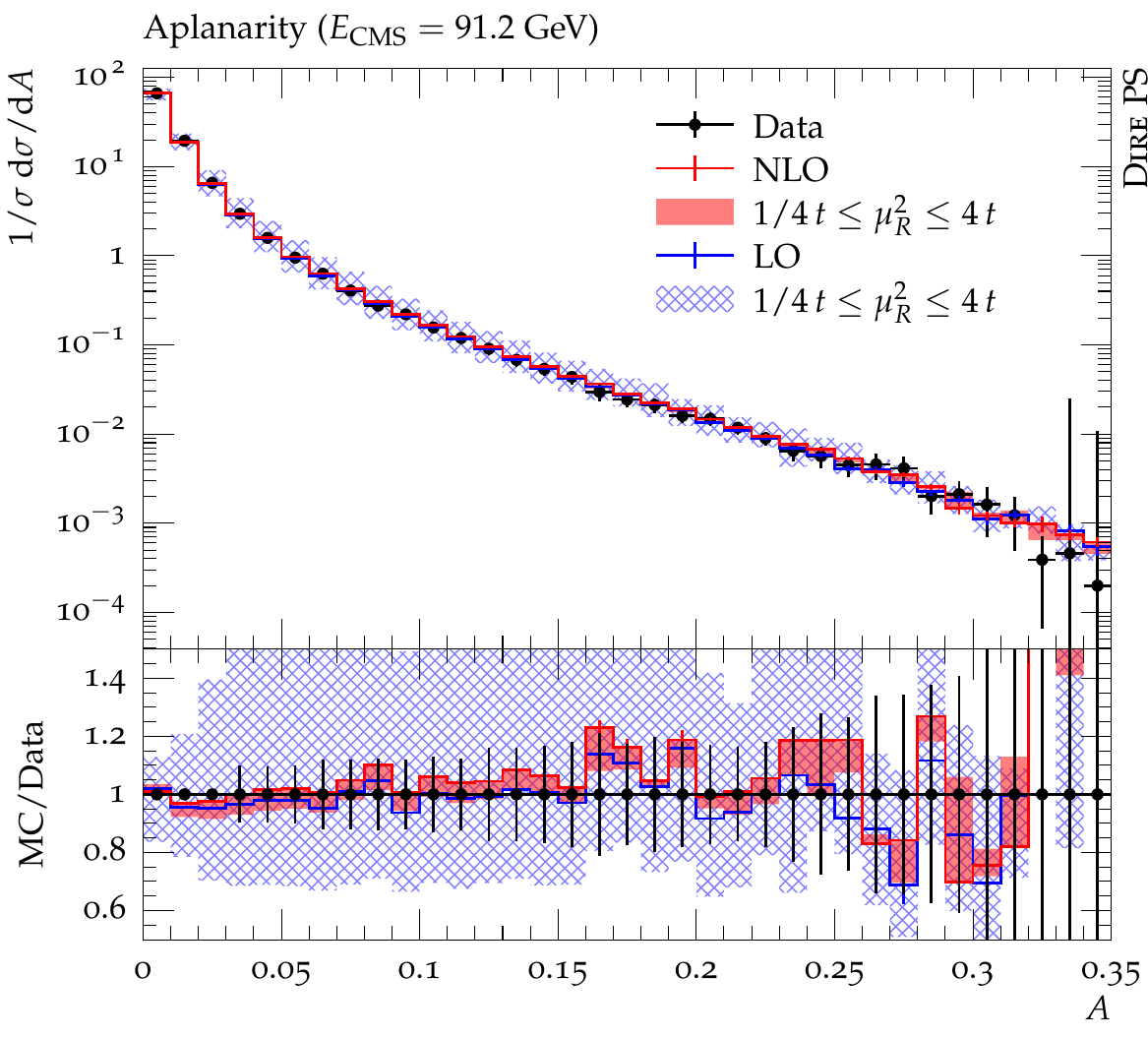}
  \caption{Results for leading and next-to-leading order DGLAP evolution
    in comparison to LEP data from~\cite{Heister:2003aj}.
    \label{fig:lep_shapes}}
\end{figure}

Figure~\ref{fig:lep_shapes} shows a comparison for event shapes measured by the
ALEPH collaboration~\cite{Heister:2003aj}. The perturbative region is to the right
of the plot, except for the thrust distribution, where it is to the left. We notice
some deviation in the predictions for jet broadening and for the $C$-parameter,
which are largely unchanged compared to the leading-order prediction. 
These deviations are mostly within the 2$\sigma$ uncertainty of the experimental
measurements, and they occur close to the nonperturbative region, which indicates
that they may be related to hadronization effects.

\begin{figure}[p]
  \centering
  \includegraphics[width=5.75cm]{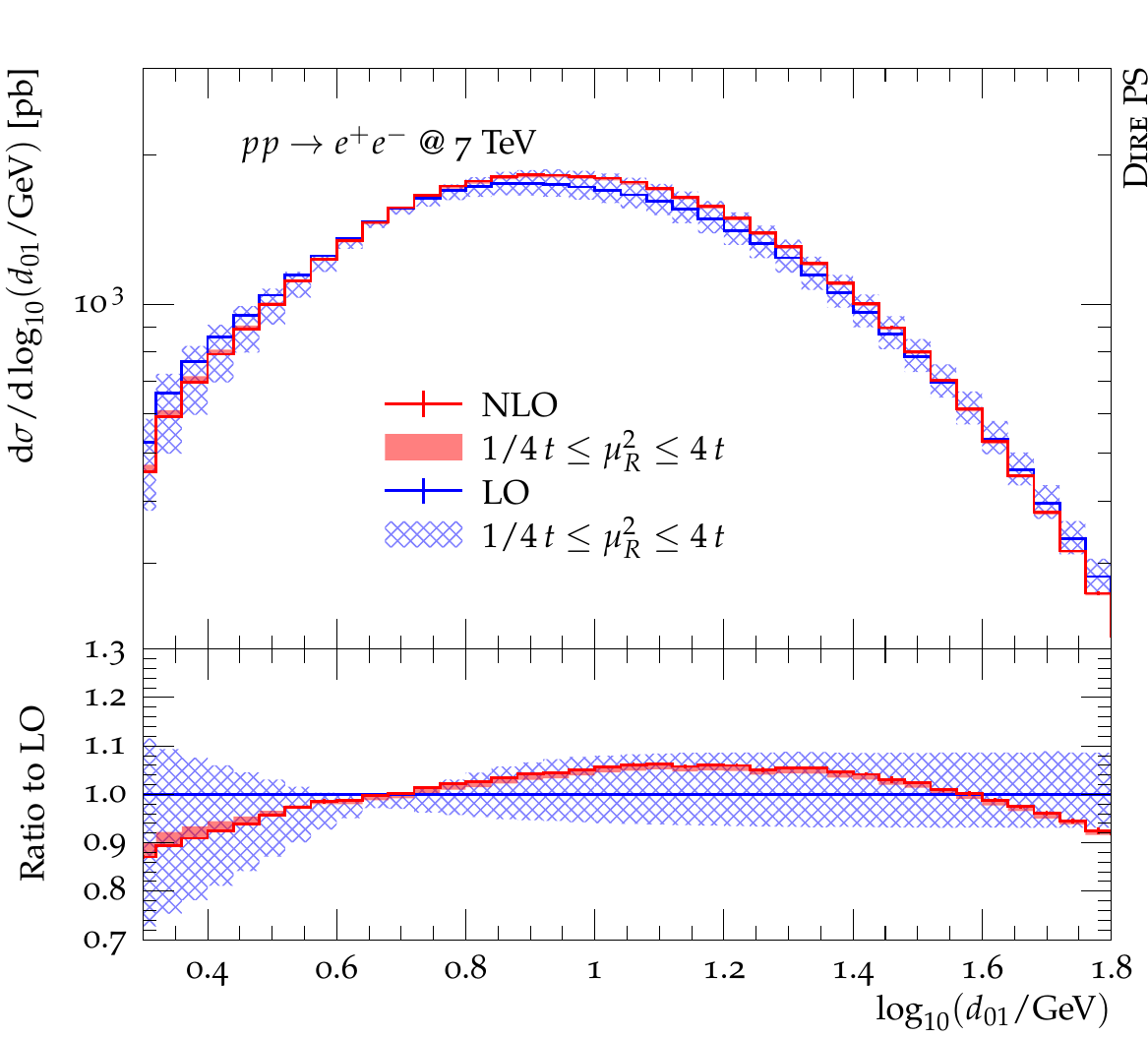}
  \includegraphics[width=5.75cm]{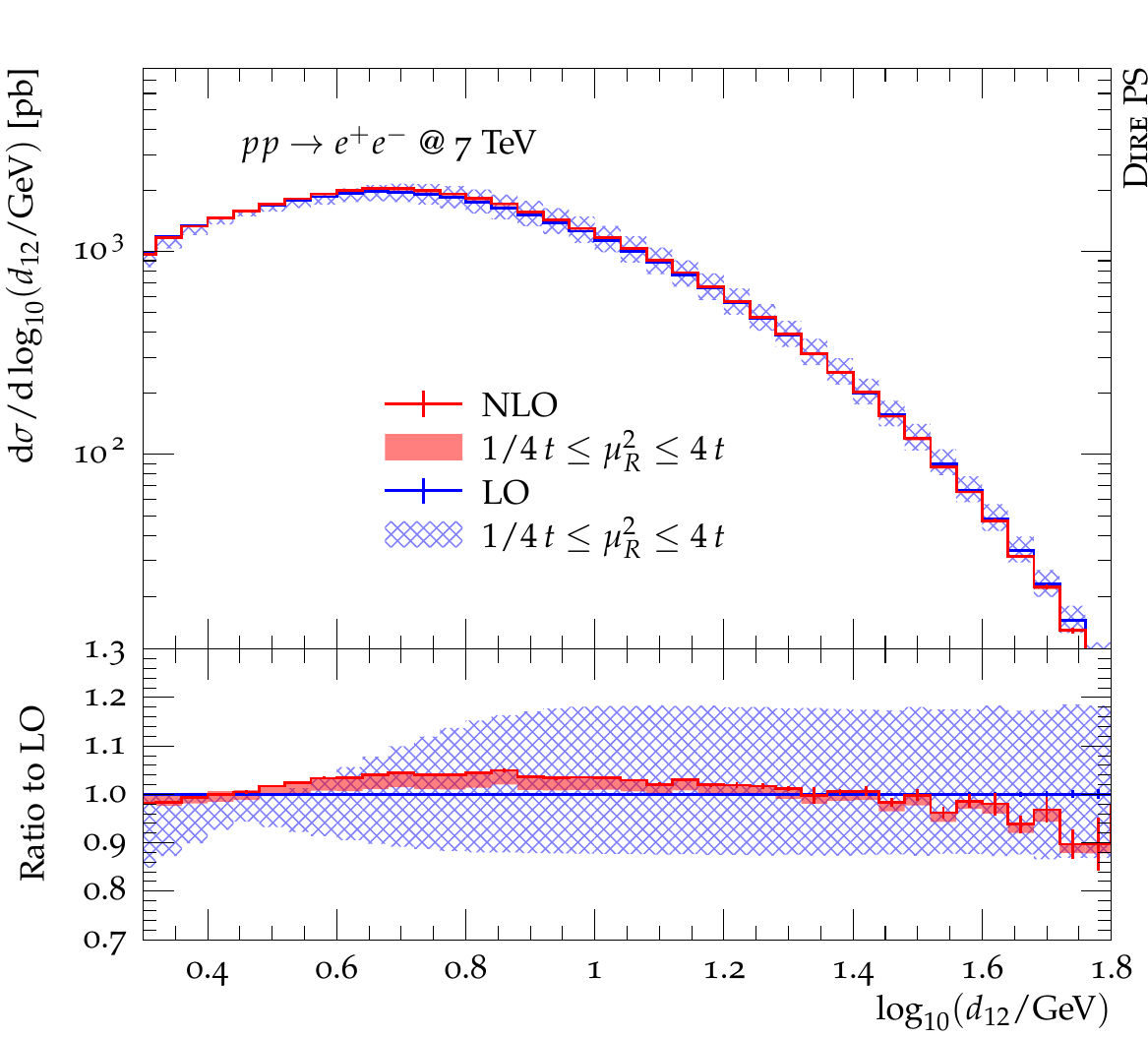}\\
  \includegraphics[width=5.75cm]{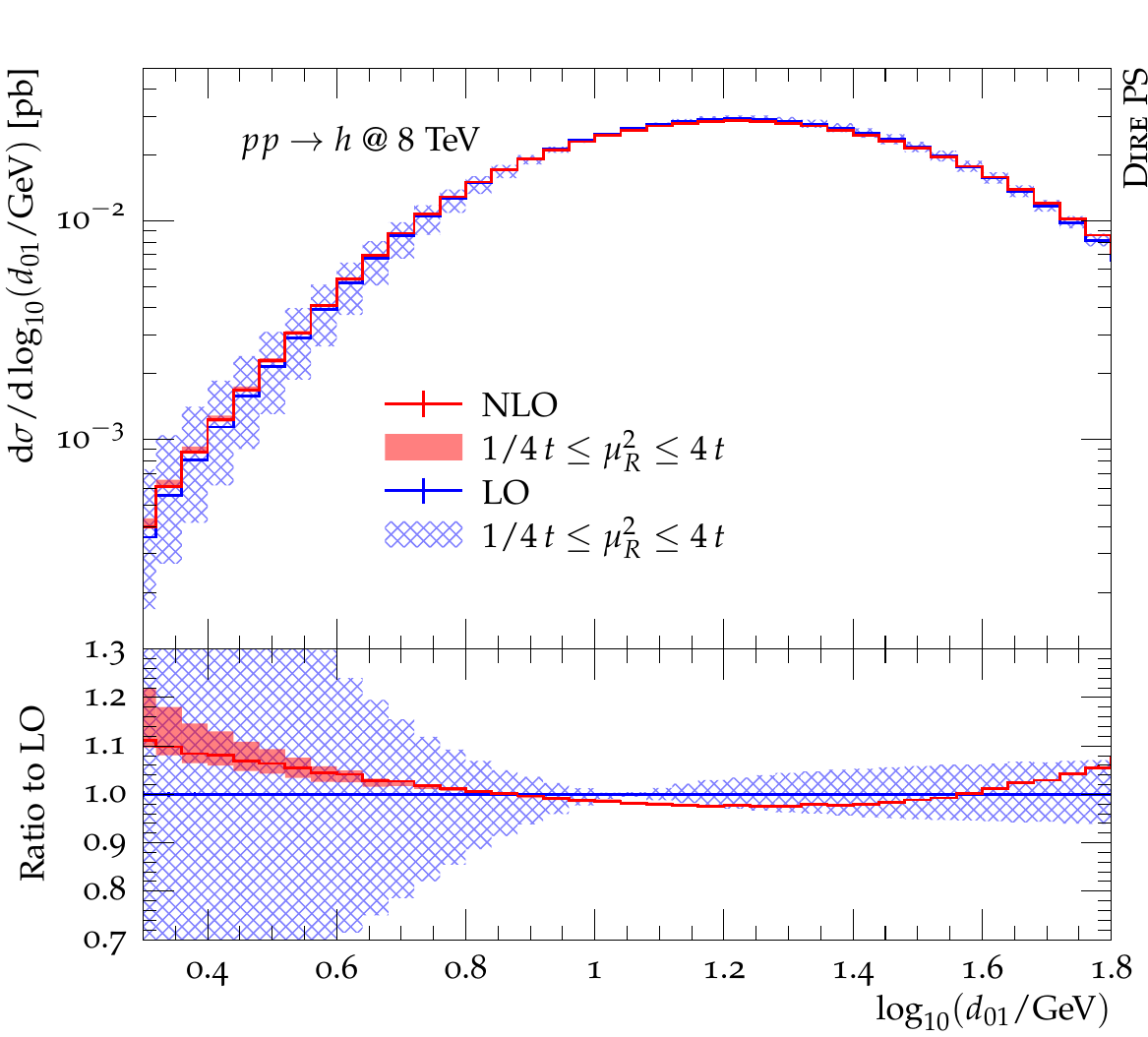}
  \includegraphics[width=5.75cm]{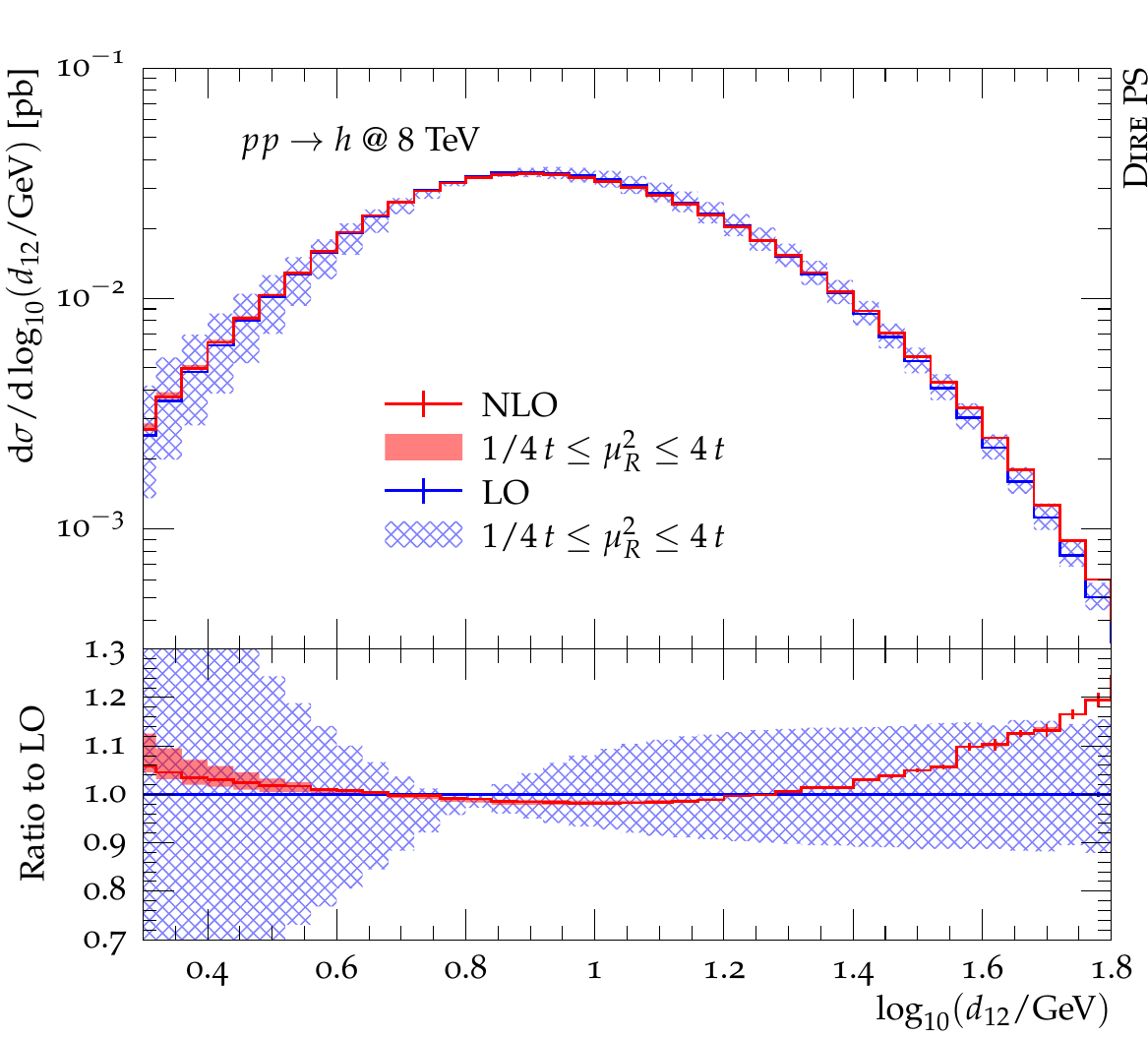}
  \caption{Predictions for leading and next-to-leading order DGLAP evolution
    for the differential $k_T$-jet resolution parameters in $pp\to e^+e^-+X$ 
    (LHC $\sqrt{s}=7$~TeV) and $pp\to h+X$ (LHC $\sqrt{s}=8$~TeV).
    \label{fig:lhc_dij}}
\end{figure}

In Figure \ref{fig:lhc_dij}, we illustrate the effect of NLO kernels on 
differential jet resolutions in Drell-Yan lepton-pair production as well as on
Higgs-boson production in gluon fusion. In both cases, the impact of varying
the renormalization scale in the parton shower is greatly reduced upon inclusion
of NLO corrections, and shape-changes of $\mathcal{O}(10\%)$ can be observed.
It is interesting to note that these shape changes have the opposite effect in
Drell-Yan lepton pair and Higgs boson production. This effect could not have been
obtained by changing the argument of $\alpha_s$ at leading order only,
as in Eq.~\eqref{eq:asct}.

\begin{figure}[p]
  \centering
  \includegraphics[width=5.75cm]{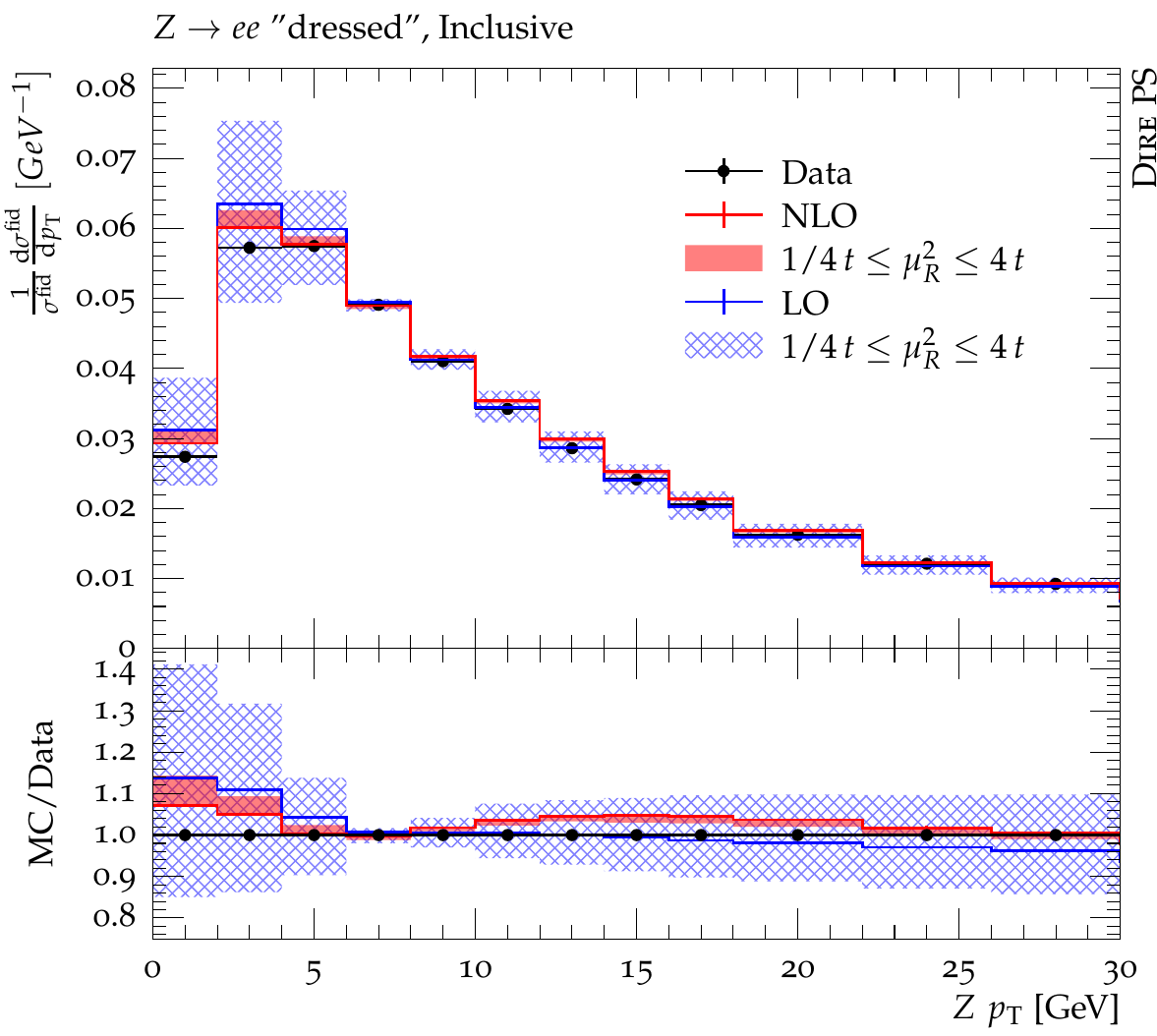}
  \includegraphics[width=5.75cm]{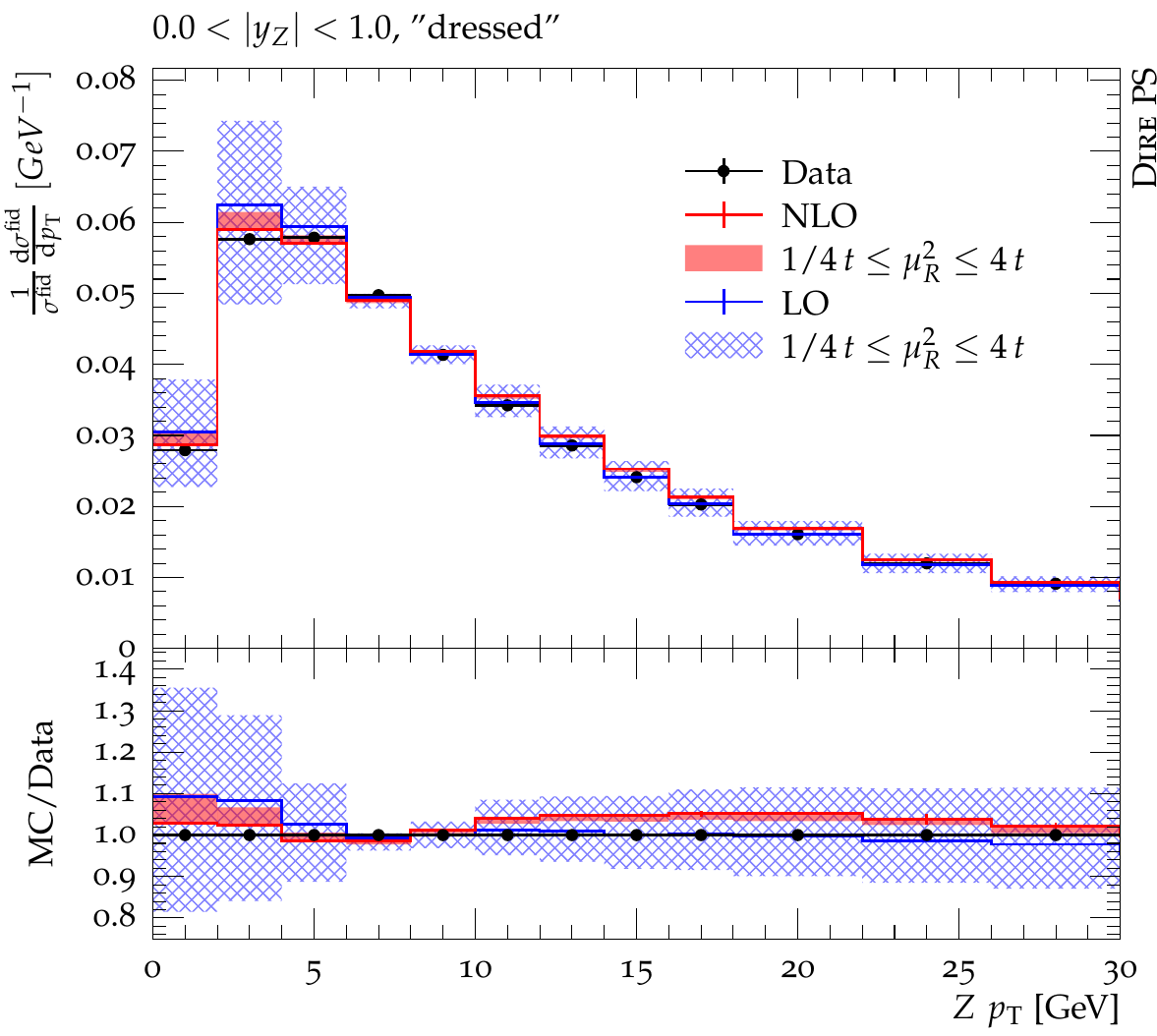}\\
  \includegraphics[width=5.75cm]{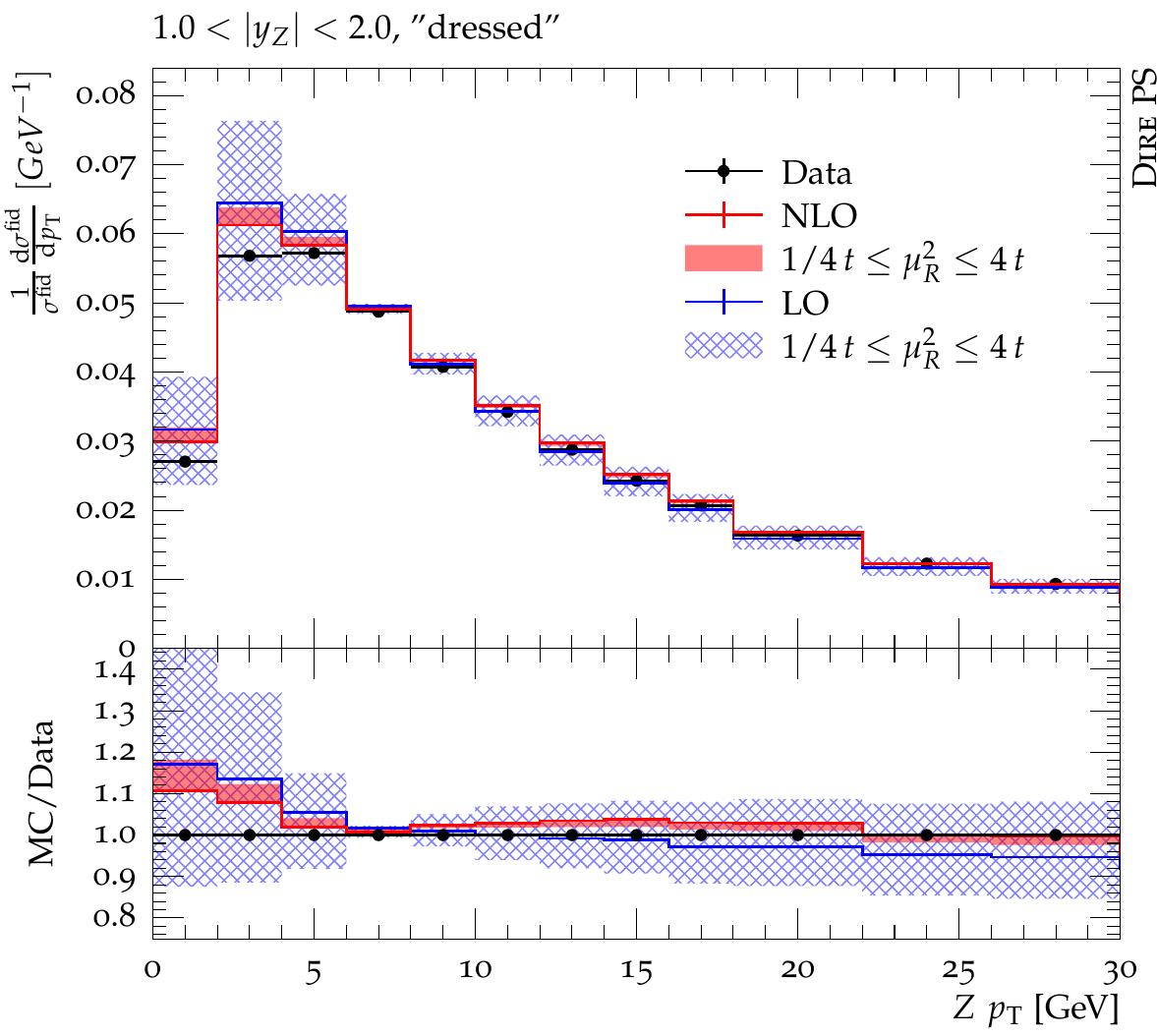}
  \includegraphics[width=5.75cm]{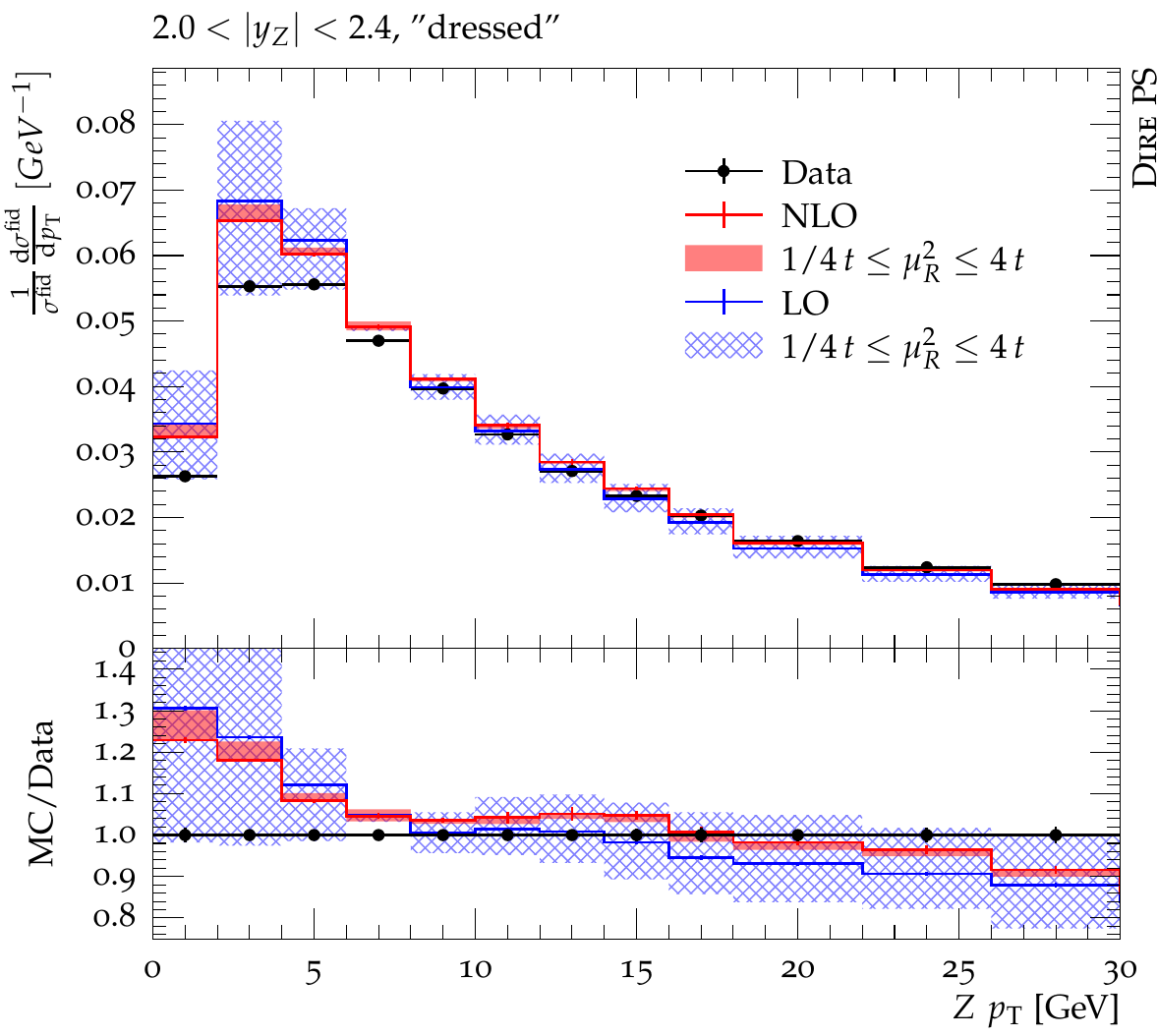}
  \caption{Results for leading and next-to-leading order DGLAP evolution
    in comparison to ATLAS data from~\cite{Aad:2014xaa}.
    \label{fig:lhc_dypt}}
\end{figure}

Figure \ref{fig:lhc_dypt} confronts \Dire with Drell-Yan transverse 
momentum spectra measured by ATLAS \cite{Aad:2014xaa}. We limit the comparison
to the soft and semi-hard region of transverse momenta, $p_T<30$~GeV. Parton shower
predictions are insufficient in the hard region, and the shower is usually
supplemented with fixed-order calculations through matching or merging in order
to improve upon this deficiency.
Note that no tuning of \Dire{}+\Pythia has been performed, neither in the
default version nor for the present publication. Our results have been obtained
with \Pythia~8.226, using the NNPDF 3.0 (NLO) PDF set \cite{Ball:2014uwa}, 
$\alpha_s(M_Z)=0.118$ throughout the simulation. The ISR/FSR shower cut-off has been
set to $3$ GeV$^2$, and a primordial transverse momentum of $k_\perp=2$ GeV was used.
All other parameters are given by the default tune of \Pythia~8 \cite{Skands:2014pea}.
NLO corrections improve the agreement with data particularly in the region
where resummation has a large impact.



\section{Summary}
\label{sec:conclusions}
In this paper we have presented an extension of the parton shower formalism
to include, for the first time, the DGLAP evolution at next-to-leading order
precision for both initial and final state radiation.  The new terms are of
order $\alpha_S^2$, and they fall into three categories: Soft terms $\propto 1/(1-z)$
which are multiplied by the two-loop cusp anomalous dimension,
and which are routinely included in parton shower simulations
through a suitable rescaling of the argument of the strong coupling.  In
addition there are genuine, non-trivial higher-order terms which
modify already existing leading order terms.  Although they are negative over
a wide region of phase space we can include them as separate terms through
existing reweighting techniques.  Finally, there are
new structures which correspond to flavor-changing transitions of the type
$q\to q'$ or $q\to\bar{q}$ and which originate in genuine $1\to 3$
transitions.  The algorithm for their simulation is detailed in a separate
publication.  Including all these terms corresponds to adding the
process-independent collinear enhanced NLO corrections present in standard
DGLAP evolution into the parton shower. The overall effect of this increased
precision is twofold.  While for $e^-e^+$ annihilations to hadrons the central
values of distributions experience only marginal shifts, the situation is
different for distributions at hadron colliders.  This is exemplified by
the transverse momentum distribution of $Z$-bosons produced at tree-level
in $q\bar{q}$ annihilation and the differential jet rates in $Z$ and Higgs
boson production through gluon fusion, respectively, which experience
some shifts of up to about 10\% relative to corresponding leading order
distributions.  In both cases, the uncertainty from variations of the
renormalization scale by factors of two is significantly reduced
when going from leading to next-to-leading order precision.  For the first
time, we are able to quote a realistic renormalization scale uncertainty
as we only add renormalization counterterms which appear at the perturbative
order to which we control the expansion of the splitting functions.

While the work presented here represents a significant improvement over
existing parton showers, it includes only parts of the higher-order
corrections.  We did not improve upon the leading color approximation
typically used in the parton shower. Ways to include such corrections have been
discussed in~\cite{Platzer:2012np}. Furthermore, we did not include the
effect of higher-order soft terms, {\it i.e.}\ the effect of multiple
unordered soft emissions.  We expect these terms to have only limited impact
on inclusive observables such as standard event shapes or the transverse
momentum of singlet particles produced at hadron colliders. They will mostly
contribute to the further stabilization of perturbative predictions for these
observables. However, we appreciate that they will certainly impact on non-global
observables such as out-of-cone radiation which in turn renders their inclusion
an important task for the future.

\begin{acknowledgments}
  \noindent
  We thank Stefano Catani, Stanislaw Jadach and Leif L{\"o}nnblad for
  numerous fruitful discussions.  This work was supported by the US
  Department of Energy under contracts DE--AC02--76SF00515 and
  DE--AC02--07CH11359,
  and by the European Commission under Grant
  Agreements PITN-GA-2012-315877 (``MCnet'') and
  PITN-GA-2012-316704 (``HiggsTools''), and by the ERC 
  Advanced Grant 340983 (``MC@NNLO'').
\end{acknowledgments}

\appendix
\section{Next-to-leading order splitting functions}
\label{app:coefficient_functions}
The components of the unregularized space-like quark splitting functions
in Eq.~\eqref{eq:ap_kernels_1_qdef} are given by
\begin{equation}\label{eq:aps_kernels_v}
  \begin{split}
    p_{qq}^{V(1)}(z)=&\;
    p^{(0)}_{qq}(z)\left[\Big(\beta_0\log z+\Gamma^{(2)}\Big)
      -2C_F\log z\left(\log(1-z)+\frac{3}{4}\right)+
      \frac{C_A}{2}\log^2 z\right]
    -\frac{4}{3}C_F T_F(1-z)\\
    &-C_F^2\,\left[\left(\frac{3}{2}+\frac{7}{2}z\right)\log z
      +\frac{1}{2}(1+z)\log^2z+5(1-z)\right]
    +C_FC_A\,\left[(1+z)\log z+\frac{20}{3}(1-z)\right]\;,\\
  \end{split}
\end{equation}
\begin{equation}\label{eq:aps_kernels_s}
    p_{qq}^{S(1)}(z)=C_FT_F\bigg[-(1+z)\log^2z
      +\left(\frac{8}{3}z^2+5z+1\right)\log z
      -\frac{56}{9}z^2+6z-2+\frac{20}{9z}\,\bigg]\;.
\end{equation}
\begin{equation}\label{eq:aps_kernels_qg}
  \begin{split}
    p_{qg}^{(1)}(z)=&\;
    p_{qg}^{(0)}(z)\bigg[\,2\beta_0\log(1-z)
      -T_F\left(\frac{4}{3}z+\frac{20}{9}\right)
      -C_F\Big(3\log(1-z)+\log^2(1-z)\Big)\bigg]\\
    &+C_F^2\,\bigg[-\frac{5}{2}-\frac{7z}{2}+\left(2+\frac{7z}{2}\right)\log z-
      \left(1-\frac{z}{2}\right)\log^2z-2z\log(1-z)\bigg]\\
    &+C_FC_A\,\bigg[\,\frac{28}{9}+\frac{65z}{18}+\frac{44z^2}{9}-
      \left(12+5z+\frac{8z^2}{3}\right)\log z+(4+z)\log^2 z+2z\log(1-z)\\
      &\qquad+S_2(z)\hat{p}^{(0)}_{qg}(-z)
      +\left(\frac{1}{2}-2\log z\log(1-z)+\frac12\log^2 z
      +\log^2(1-z)-\frac{\pi^2}{6}\right)\,\hat{p}^{(0)}_{qg}(z)
      \bigg]\;.
  \end{split}
\end{equation}
We use the auxiliary function $S_2$ defined in~\cite{Ellis:1991qj}
\begin{equation}
  \begin{split}
    S_2(z)=\;&-2\,\Li_2\Big(\frac{1}{1+z}\Big)
    +\frac{1}{2}\log^2 z-\log^2(1-z)+\frac{\pi^2}{6}\;.
  \end{split}
\end{equation}
The unregularized gluon splitting functions at NLO are given by
\begin{equation}\label{eq:aps_kernels_gq}
  \begin{split}
    p_{gq}^{(1)}(z)=&\;
    C_F T_F\bigg[\,
      p_{gq}^{(0)}(z)\,\bigg(2\log^2\left(\frac{1-z}{z}\right)-
      4\log\left(\frac{1-z}{z}\right)-\frac{2\pi^2}{3}+10\bigg)
      \\
      &\quad\quad+4-9z-(1-4z)\log z-(1-2z)\log^2 z+4\log(1-z)
      \bigg]+C_AS_2(z)p_{gq}^{(0)}(-z)
    \\
    &+C_A T_F\bigg[\,
      p_{gq}^{(0)}(z)\bigg(\,
      -\log^2z+\frac{44}{3}\log z-2\log^2(1-z)+4\log(1-z)+\frac{\pi^2}{3}
      -\frac{218}{9}
      \bigg)\\
      &\quad\quad+\frac{182}{9}+\frac{14z}{9}+\frac{40}{9z}+
      \left(\frac{136z}{3}-\frac{38}{3}\right)\log z-
      4\log(1-z)-(2+8z)\log^2 z
      \bigg]\;.
  \end{split}
\end{equation}
\begin{equation}\label{eq:aps_kernels_gg}
  \begin{split}
    p_{gg}^{(1)}(z)=&\;
    p^{(0)}_{gg}(z)\,\bigg[\,\Gamma^{(2)}+
      C_A\bigg(-2\log z\log(1-z)+\frac{1}{2}\log^2 z\bigg)\bigg]\,+\,
    C_AS_2(z)p^{(0)}_{gg}(-z)\\
    &+C_F T_F\,\bigg[\,-16+8z+\frac{20z^2}{3}+\frac{4}{3z}-
      (6+10z)\log z-(2+2z)\log^2z\bigg]\\
    &+C_A T_F\bigg[\,2-2z+\frac{26}{9}\left(z^2-\frac{1}{z}\right)
      -\frac{4}{3}(1+z)\log z\bigg]\\
    &+C_A^2\bigg[\,\frac{27}{2}(1-z)+\frac{67}{9}\left(z^2-\frac{1}{z}\right)
      -\left(\frac{25}{3}-\frac{11z}{3}+\frac{44z^2}{3}\right)\log z
      +4(1+z)\log^2z
      \bigg]\,.
  \end{split}
\end{equation}
The components of the unregularized time-like quark splitting functions
in Eq.~\eqref{eq:ap_kernels_1_qdef} are given by
\begin{equation}\label{eq:apt_kernels_v}
  \begin{split}
    p_{qq}^{V(1)}(z)=&\;
    p^{(0)}_{qq}(z)\left[\Big(\beta_0\log z+\Gamma^{(2)}\Big)
      +2C_F\log z\left(\log\frac{1-z}{z}+\frac{3}{4}\right)+
      \frac{C_A}{2}\log^2 z\right]-\frac{4}{3}C_F T_F(1-z)\\
    &-C_F^2\,\left[\left(\frac{7}{2}+\frac{3}{2}z\right)\log z
      -\frac{1}{2}(1+z)\log^2z+5(1-z)\right]
    +C_FC_A\,\left[(1+z)\log z+\frac{20}{3}(1-z)\right]\;,\\
  \end{split}
\end{equation}
\begin{equation}\label{eq:apt_kernels_s}
  p_{qq}^{S(1)}(z)=\;C_FT_F\left[\,(1+x)\log^2 x
  -\left(\frac{8}{3}z^2+9z+5\right)\log z
  +\frac{56}{9}z^2+4z-8-\frac{20}{9z}\,\right]\;.
\end{equation}
\begin{equation}\label{eq:apt_kernels_qg}
  \begin{split}
    p_{qg}^{(1)}(z)=&\;
    C_F^2\,\bigg[\,\hat{p}_{qg}^{(0)}(z)\bigg(\,
      \log^2(1-z)+4\log z\log(1-z)-8S_1(z)-\frac{4}{3}\pi^2\bigg)\\
      &\quad\quad-\frac12+\frac{9z}{2}-\left(8-\frac{z}{2}\right)\log z+
      2z\log(1-z)+\left(1-\frac{z}{2}\right)\log^2z
      \bigg]+C_AS_2(z)p_{qg}^{(0)}(-z)\\
    &+C_FC_A\,\bigg[\,\frac{62}{9}-\frac{35z}{18}-\frac{44z^2}{9}+
      \left(2+12z+\frac{8}{3}z^2\right)\log z-2z\log(1-z)-(4+z)\log^2z\\
      &\qquad-
      \hat{p}_{qg}^{(0)}(z)\bigg(2\log z\log(1-z)+3\log z+\frac32\log^2z
      +\log^2(1-z)-8S_1(z)-\frac{7\pi^2}{6}-\frac{17}{18}\bigg)
      \bigg]\;.
  \end{split}
\end{equation}
We use the auxiliary function $S_1$ defined in~\cite{Ellis:1991qj}
\begin{equation}
  \begin{split}
    S_1(z)=\;&\Li_2(z)+\log z\log(1-z)-\frac{\pi^2}{6}\;.
  \end{split}
\end{equation}
The unregularized gluon splitting functions at NLO are given by
\begin{equation}\label{eq:apt_kernels_gq}
  \begin{split}
    p_{gq}^{(1)}(z)=&\;
    T^2_F\bigg[\,-\frac{8}{3}-p_{gq}^{(0)}(z)
      \left(\frac{16}{9}+\frac{8}{3}\log z+\frac{8}{3}\log(1-z)\right)\bigg]\\
    &+C_F T_F\bigg[\,
      p_{gq}^{(0)}(z)\,
      \left(-2\log^2(z(1-z))-2\log\left(\frac{1-z}{z}\right)+
      16S_1(z)+2\pi^2-10\right)\\
      &\qquad-2+3z-(7-8z)\log z-4\log(1-z)+(1-2z)\log^2z
      \bigg]+C_A S_2(z)p_{gq}^{(0)}(-z)\\
    &+C_A T_F\bigg[\,
      -\frac{152}{9}+\frac{166z}{9}-\frac{40}{9z}-
      \left(\frac{76z}{3}+\frac{4}{3}\right)\log z+
      4\log(1-z)+(2+8z)\log^2 z\\
      &\qquad+\hat{p}_{gq}^{(0)}(z)\bigg(\,
      8\log z\log(1-z)-\log^2z-\frac{4}{3}\log z+\frac{10}{3}\log(1-z)\\
      &\qquad\qquad+2\log^2(1-z)-16S_1(z)-\frac{7\pi^2}{3}+\frac{178}{9}
      \bigg)\bigg]\\
  \end{split}
\end{equation}
\begin{equation}\label{eq:apt_kernels_gg}
  \begin{split}
    p_{gg}^{(1)}(z)=&\;
    p^{(0)}_{gg}(z)\,\bigg[2\beta_0\log z+\Gamma^{(2)}+
      \,C_A\log z\,\bigg(2\log(1-z)-\frac{3}{2}\log^2z\bigg)\bigg]\,+\,
    C_AS_2(z)p^{(0)}_{gg}(-z)\\
    &+C_F T_F\,\bigg[\,-4+12z-\frac{164}{9}z^2
      \left(10+14z+\frac{16z^2}{3}+\frac{16}{3z}\right)\log z
      +\frac{92}{9z}+2(1+z)\log^2z\bigg]\\
    &+C_A T_F\bigg[\,2-2z+\frac{26}{9}\left(z^2-\frac{1}{z}\right)
      -\frac{4}{3}(1+z)\log z\bigg]\\
    &+C_A^2\bigg[\,\frac{27}{2}(1-z)+\frac{67}{9}\left(z^2-\frac{1}{z}\right)
      +\left(\frac{11}{3}-\frac{25}{3}z+\frac{44}{3z}\right)\log z
      -4(1+z)\log^2z
      \bigg]\,.
  \end{split}
\end{equation}

\bibliography{journal}
\end{document}